%% file: tdfinder.tex
\documentclass{LMCS}

\def\dOi{11(3:15)2015}
\lmcsheading%
{\dOi}
{1--27}
{}
{}
{Jan.~13, 2014}
{Sep.~17, 2015}
{}

\ACMCCS{[{\bf Software and its engineering}]: Software organization
  and properties---Software functional properties---Formal
  methods---Software verification; [{\bf Theory of computation}]:
  Semantics and reasoning---Program reasoning---Program verification}

\usepackage{hyperref}

\usepackage[usenames,dvipsnames,svgnames,table]{xcolor}

\usepackage{tikz}
\usetikzlibrary{arrows,calc,automata,petri,shapes,positioning,shadows,decorations.markings,decorations.pathreplacing,fit,matrix}

\tikzset{
>=stealth',
help lines/.style={dashed, thick},
axis/.style={<->},
important line/.style={thick},
connection/.style={thick, dotted},
}

\usepackage{amsfonts, stmaryrd, amssymb, amsthm, amsmath}

\usepackage[inference]{semantic}

\usepackage{bm,xspace,multirow}

\input{commonCmds}

\input{tikzEx.tex}

\theoremstyle{plain}
\theoremstyle{plain}

\begin{document}

\title[Compositional Verification for Timed Systems]
{Compositional Verification for Timed Systems\\ based on Automatic Invariant Generation\rsuper*}

\author[S.~Ben Rayana]{Souha Ben Rayana\rsuper a}
\address{{\lsuper{a--d}}Univ. Grenoble Alpes, VERIMAG, F-38000 Grenoble}
\email{\{Souha.BenRayana,lastefan,Saddek.Bensalem\}@imag.fr}

\author[L.~A\c stef\u anoaei]{L\u acr\u amioara ~A\c stef\u
  anoaei\rsuper b}
\address{\vspace{-18 pt}}

\author[S.~Bensalem]{Saddek Bensalem\rsuper c}
\address{\vspace{-18 pt}}

\author[M.~Bozga]{Marius Bozga\rsuper d}
\address{{\lsuper{c,d}}CNRS, VERIMAG, F-38000 Grenoble, France}
\email{\{Marius.Bozga,Jacques.Combaz\}@imag.fr}

\author[J.~Combaz]{Jacques Combaz}
\address{\vspace{-18 pt}}


\titlecomment{{\lsuper*}Research supported by the European Integrated Project 257414
  ASCENS and ICT Collaborative Project 288175 CERTAINTY}

\begin{abstract}
  We propose a method for compositional verification to address the
  state space explosion problem inherent to model-checking timed
  systems with a large number of components. The main challenge is to
  obtain pertinent global timing constraints from the timings in the
  components alone. To this end, we make use of auxiliary clocks to
  automatically generate new invariants which capture the constraints
  induced by the synchronisations between components. The method has
  been implemented in the RTD-Finder tool and  successfully experimented on several benchmarks.
\end{abstract}

\maketitle

\section{Introduction}
\label{sec:intro}

Compositional methods in verification have been developed to cope with
state space explosion. Generally based on divide et impera principles,
these methods attempt to break monolithic verification problems into
smaller sub-problems by exploiting either the structure of the system
or the property or both. Compositional reasoning can be used in
different manners e.g., for deductive verification, assume-guarantee,
contract-based verification, compositional generation, etc.

The development of compositional verification for timed systems
remains however challenging. State-of-the-art tools
\cite{uppaal,kronos,red06,Romeo2005} for the verification of such
systems are mostly based on symbolic state space exploration, using
efficient data structures and particularly involved exploration
techniques.  In the timed context, the use of compositional reasoning
is inherently difficult due to the synchronous model of time. Time
progress is an action that synchronises continuously all the
components of the system. Getting rid of the time synchronisation is
necessary for analysing independently different parts of the system
(or of the property) but becomes problematic when attempting to
re-compose the partial verification results.  Nonetheless,
compositional verification is actively investigated and several
approaches have been recently developed and employed in timed
interfaces \cite{Alfaro02} and contract-based assume-guarantee
reasoning \cite{Ecdar,AutomaticCompERAs}.

In this paper, we propose a different approach for exploiting
compositionality for analysis of timed systems. The driving principle
is to use invariants as approximations to exact reachability analysis,
the default technique in model-checking.  We show that rather
precise invariants can be computed compositionally, from the separate
analysis of the components in the system and from their composition
glue. This method is proved to be sound for the verification of safety
state properties. However, it is not complete.

The starting point is the verification method of~\cite{dfinder},
summarised in Figure~\ref{fig:VR}. The method exploits
compositionality as explained next. Consider a system consisting of
components $\cn_i$ interacting by means of a set $\gamma$ of
multi-party interactions, and let $\varphi$ be a system property of
interest. Assume that all $\cn_i$ as well as the composition through
$\gamma$ can be independently characterised by means of component
invariants $\ic(B_i)$, respectively interaction invariant
$\iim(\gamma)$. The connection between the invariants and the system
property $\varphi$ can be intuitively understood as follows: if $\varphi$
can be proved to be a logical consequence of the conjunction of
components and interaction invariants, then $\varphi$ holds for the
system.

\begin{figure}[htp]
\begin{center}
\begin{tabular}{c}
  \inference{\vdash \big(\bigwedge_i \ic(\cn_i)\big) \wedge \iim(\gamma) \rightarrow \varphi}{ \|_{\gamma}\cn_i \models \Box\, \varphi } [\vr]
\end{tabular}
\caption{Compositional verification}\label{fig:VR}
\end{center}
\end{figure}

In the rule $\vr$\; the symbol `` $\vdash$ '' is used to underline
that the logical implication can be effectively proved (for instance
with an SMT solver) and the notation ``$\|_{\gamma}\cn_i \models \Box\, \varphi$'' is to
be read as ``$\varphi$ holds in every reachable state of $\|_{\gamma}\cn_i$''.

The verification rule (VR) in \cite{dfinder} has been developed for untimed
systems.  Its direct application to timed systems may be weak as
interaction invariants do not capture global timings of interactions
between components. The key contribution of this paper is to improve the
invariant generation method so to better track such global timings by means
of auxiliary \textit{history clocks} for actions and interactions.  At
component level, history clocks expose the local timing constraints
relevant to the interactions of the participating components. At
composition level, extra constraints on history clocks are enforced due to
the simultaneity of interactions and to the synchrony of time progress.

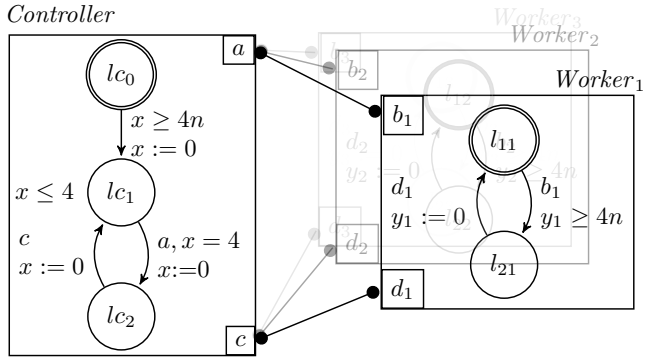
\begin{figure}[htp]
\begin{center}
\input{abstractExTikz.tex}
\end{center}
\caption{A timed system}
\label{fig:ncw}
\end{figure}

As an illustration, let us consider as running example the timed
system in Figure~\ref{fig:ncw} which depicts a ``controller''
component serving $n$ ``worker'' components, one at a time. The
interactions between the controller and the workers are defined by the
set of synchronisations $\{ (a\mid b_i), (c\mid d_i) \mid i \leq n
\}$.  Periodically, after every 4 units of time, the controller
synchronises its action $a$ with the action $b_i$ of any worker $i$
whose clock shows at least $4n$ units of time.  Initially, such a
worker exists because the controller waits for $4n$ units of time
before interacting with workers.  The cycle repeats forever because
there is always a worker ``willing'' to do $b$, that is, the system is
deadlock-free.  Proving deadlock-freedom of the system requires to
establish that when the controller is at location $lc_1$ there is at
least one worker such that $y_i - x \geq 4n - 4$.  Unfortunately, this
property cannot be shown if we use (VR) as it is in~\cite{dfinder}.
Intuitively, this is because the proposed invariants are too weak to
infer cross constraints relating the clocks of the controller and
those of the workers: interaction invariants $\iim(\gamma)$ relates
only locations of components and thus at most eliminates unreachable
configurations like $(lc_1, \dots, l_{2i},\dots)$, while the component
invariants can only state local conditions on clocks such as $x \leq
4$ at $lc_1$.  Using history clocks allows to recover additional
constraints. For example, after the controller returns from $lc_2$ to
$lc_1$ for the first time, whenever it reaches $lc_1$ again, there
exists a worker $i$ whose clock has an equal value as that of the
controller. Similarly, history clocks allow to infer that different
$(a\mid b_i)$ interactions are separated by at least 4 time units.
These constraints altogether are sufficient to prove the deadlock
freedom property.

\subsection*{Organisation of the paper}  
This paper is essentially an extended version of the conference paper
\cite{abbbc14}. The extension is threefold with respect to (1)
incorporating proofs, (2) detailing technicalities about handling
initial states, and (3) formalising three heuristics to speed up and
simplify invariant generation. Section~\ref{sec:frtbip} recalls the
needed definitions for modelling timed systems and their
properties. Section~\ref{s:bm} presents our method for compositional
generation of invariants. Section~\ref{sec:heuristics} describes the
heuristics while Section~\ref{sec:impl} shows their use in the case
studies we experimented with in our
implementation. Section~\ref{sec:conc} concludes.


\section{Timed Systems and Properties}
\label{sec:frtbip}
In the framework of the present paper, components are timed automata and
systems are compositions of timed automata with respect to multi-party
interactions. The timed automata we use are essentially the ones from
\cite{alur94}, however, slightly adapted to embrace a uniform notation
throughout the paper.

\begin{defi}[Syntax]
  A component is a timed automaton $(L, A, \X, T, \inv, s_0)$ where
  $L$ is a finite set of locations, $A$ a finite set of actions, $\X$
  is a finite set of local\footnote{Locality is essential for avoiding
    side effects which would break compositionality and local
    analysis.}  clocks, $T \subseteq L\times(A\times \mathcal C\times
  2^{\X})\times L$ is a set of edges labelled with an action, a guard,
  and a set of clocks to be reset, $\inv: L\rightarrow {\mathcal C}$
  assigns a time progress condition\footnote{To avoid confusion with
    invariant properties, we prefer to adopt the terminology of ``time
    progress condition'' from \cite{bornot98} instead of ``location
    invariants''.} to each location.  $\mathcal C$ is the set of clock
  constraints and $s_0 \in L \times {\mathcal C}$ provides the initial
  configuration. A clock constraint is defined by the grammar:
  \[ C ::= \true \mid x\#\mathit{ct} \mid x - y \#\mathit{ct}
  \mid C \wedge C \] with $x, y\in \X$, $\#\in\{<,\le,=,\ge,>\}$ and
  $\mathit{ct} \in \mathbb{Z}$.  Time progress conditions are restricted to
  conjunctions of constraints as $x \leq \mathit{ct}$.
\end{defi}

Before recalling the semantics of a component, we first fix some notation.
Let $\mathbf{V}$ be the set of all clock valuation functions $\bv:
\X\rightarrow \mathbb{R}_{\ge 0}$.  For a clock constraint $C$, $\bv \models C$
denotes the evaluation of $C$ in $\bv$. The notation $\bv + \delta$
represents a new $\mathbf{v'}$ defined as $\mathbf{v'}(x) = \bv(x) +
\delta$ while $\bv[r]$ represents a new $\mathbf{v'}$ which assigns any $x$
in $r$ to 0 and otherwise preserves the values from $\mathbf{v}$.

\begin{defi}[Semantics]
  The semantics of a component $B = (L, A, \X, T, \inv, s_0)$ is given by
  the labelled transition system $(Q, A, \rightarrow, Q_0)$ where $Q
  \subseteq L\times \mathbf{V}$ denotes the states of $B$,
  ${\rightarrow} \subseteq Q\times (A\cup \mathbb{R}_{\ge 0}) \times
  Q$ denotes the transitions according to the rules:
    \begin{itemize}
    \item $(l,\bv) \transit{\delta} (l,\bv+\delta)$ if $ \big(\forall \delta'\in[0,\delta]\big).(\inv(l)(\bv+\delta'))$ (time progress);
    \item $(l,\bv) \transit{a} (l',\bv[r])$ if $\big(l,(a,g,r),l'\big)\in T$, $g(\bv)\wedge \inv(l')(\bv[r])$ (action step).
    \end{itemize}
    and $Q_0 = \{ (l_0, \bv_0) | s_0 = (l_0, c_0) \wedge c_0(\bv_0) \}$ denotes the initial states. 
\end{defi}

Because the semantics defined above is in general infinite, we work with
the so called zone graph \cite{henzinger94} as a finite symbolic
representation. The symbolic states in a zone graph are pairs $(l, \zeta)$
where $l$ is a location of $\cn$ and $\zeta$ is a \textit{zone}, a set of
clock valuations defined by clock constraints.  The initial configuration
$s_0=(l_0,c_0)$ corresponds trivially to a symbolic state $(l_0, \zeta_0)$.
Given a symbolic state $(l,\zeta)$, its successor with respect to a
transition $t$ of $\cn$ is denoted as $\msucc(t, (l,\zeta))$ and defined by
means of its timed and its discrete successor:
\begin{itemize}
\item $\tsucc((l,\zeta)) = (l,\nearrow\zeta \cap \inv(l))$
\item $\dsucc(t, (l,\zeta)) = (l',(\zeta \cap g)[r]\cap \inv(l'))$ if $t = \big(l, (\_, g, r), l'\big)$
\item $\msucc(t, (l,\zeta)) = \close(\tsucc(\dsucc(t, (l, \zeta))))$
\end{itemize}
where $\nearrow, [r], \close$ are usual operations on zones: $\nearrow
\zeta$ is the forward diagonal projection of $\zeta$, i.e., it
contains any valuation $\mathbf{v'}$ for which there exists a real
$\delta$ such that $\mathbf{v'}-\delta$ is in $\zeta$; $\zeta[r]$ is
the set of all valuations in $\zeta$ after applying the resets in $r$;
$\close(\zeta)$ corresponds to normalising $\zeta$ such that all
bounds on clocks and clock differences are either bounded by some
finite value or infinite. Since our use of invariants is only as
over-approximations of the reachable states, a more thorough
discussion on normalisation is not relevant for the present paper. The
interested reader may refer to \cite{bengtssonY03,bouyer04} for more
precise definitions.

A symbolic execution of $\cn$ is a sequence of
symbolic states $s_0, \dots, s_i, \dots$\footnote{We tacitly assume that
  $s_0$ is such that $s_0=\tsucc{(s_0)}$.  If this is not the case, one can
  always consider $\tsucc{(s_0)}$ instead of $s_0$ for the definition of symbolic
  executions and reachable states.} such that for any $i> 0$, there exists
a transition $t$ for which $s_i$ is $\msucc(t, s_{i-1})$. The set of
reachable symbolic states of $\cn$ is $\reach_B(s_0)$
where $\reach_B$ is defined recursively as:
\begin{align*}
 & \reach_B(s) = \{s\} \cup \displaystyle{\bigcup_{t\in T}} \reach_B(\msucc(t, s))
\end{align*}
for an arbitrary $s$ and $T$ the set of transitions in $\cn$.  We remind
that the set $\reach_B(s_0)$ can be shown finite knowing that the number
of normalised zones is finite.  In general, the symbolic zone graph
provides an over-approximation of the set of reachable states.  This
over-approximation is exact only for timed automata without diagonal
constraints \cite{bengtssonY03, bouyer04}.

In our framework, components communicate by means of
\emph{interactions}, which are synchronisations between actions.
Given $n$ components $(\cn_i)_{i=1,\dots,n}$, with disjoint sets of actions
$A_i$, an interaction is a subset $\alpha \subseteq \cup_i A_i$
containing at most one action per component.  We denote interactions
$\alpha$ as sets $\{ a_i \}_{i \in I}$, with $a_i \in A_i$ for all $i
\in I \subseteq \{ 1, \ldots, n \}$. For readability, in examples, we
use the alternative notation $(a_1 \mid a_2 \mid \dots \mid a_i)$
instead. Given a set of interactions $\gamma$, we denote by
$Act(\gamma)$ the set of actions involved in $\gamma$, that is,
$Act(\gamma) = \cup_{\alpha \in \gamma} \alpha$.

\begin{defi}[Timed System] 
For a given $n$ and $i \in \{1,\dots, n\}$ let $\cn_i$ = $(L_i$,
$A_i$, $\X_i$, $T_i$, $\inv_i$, $s_{0i})$ be $n$ components with
disjoint sets of actions and initial states $s_{0i} = (l_{0i},
c_{0i})$.  Let $\gamma$ be a set of interactions constructed from
$\cup_i A_i$.  The \emph{timed system} $\|_\gamma\cn_i$ is defined as
the component $(L,\gamma,\X,T_{\gamma},\inv,s_0)$ where $L = \times_i
L_i$, $\X = \cup_i \X_i$, $\inv(\bar{l})=\bigwedge_i\inv(l_i)$,
$s_0=((l_{01}, ..., l_{0n}), \bigwedge_i c_{0i})$ and
  $$T_\gamma = \left\{ (\bar{l}, (\alpha,g,r), \bar{l}') ~~~ \begin{array}{| l} 
      ~~ \bar{l} = (l_1,..., l_n) \in L, ~~\bar{l}' = (l'_1, ..., l'_n) \in L \\
      ~~ \alpha = \{a_i\}_{i\in I} \in \gamma,
      ~~ \forall i\in I. (l_i, (a_i,g_i,r_i), l'_i) \in T_i, 
      ~~ \forall i\not\in I. l_i = l'_i \\ 
      ~~ ~~ ~~ g = \bigwedge_{i\in I}g_i, ~~r = \bigcup_{i\in I} r_i
  \end{array}\right\}$$
\end{defi}

In the timed system $\|_\gamma \cn_i$, a component $\cn_i$ can execute
an action $a_i$ only as part of an interaction $\alpha$, $a_i \in
\alpha$, that is, along with the execution of all other actions $a_j
\in \alpha$\footnote{To simplify the notation, we omit unary interactions and the
  actions for transitions involved in them. For example, in
  Figure~\ref{fig:ncw}, the initial transition in $\ctn$ does not have
  an explicit action associated.}. This corresponds to the usual notion of multi-party
interaction. We note that interactions can only restrict the behaviour
of components, i.e., the states reached by $\cn_i$ in $\|_\gamma
\cn_i$ belong to $\reach_{B_i}( s_{0i})$. This is a property which is
exploited in the verification rule (VR) in Figure~\ref{fig:VR}. 

To give a logical characterisation of components and their properties, we
use invariants. An invariant $\Phi$ is a state predicate which holds in
every reachable state of $\cn$, in symbols, $B \models \Box \Phi$.  We use
$\ic(\cn)$ and $\iim(\gamma)$, to denote \textbf{component}, respectively
\textbf{interaction invariants}.  For component invariants, our choice is
to work with their reachable symbolic set. More precisely, for component
$\cn$, its associated component invariant $\ic(\cn)$ is the disjunction
of $(l \wedge \zeta)$ for all symbolic states $(l, \zeta)$ in
$\reach_{B}(s_{0})$.  To ease the reading, we abuse of notation and use
$l$ as a place holder for a state predicate ``$\mathit{at}(l)$'' which
holds in any symbolic state with location $l$, that is, the semantics of
$\mathit{at}(l)$ is given by $(l,\zeta) \models \mathit{at}(l)$.  As an
example, the component invariants for the example in Figure~\ref{fig:ncw}
with one worker are:
\begin{align*}
  \ic( \ctn ) & = (lc_0 \wedge x \geq 0) \vee (lc_1 \wedge 4 \geq x \geq 0) \vee (lc_2 \wedge x \geq 0)\\
  \ic( \cwkO )& = (l_{11} \wedge y_1 \geq 0) \vee (l_{21} \wedge y_1 \geq 4).
\end{align*}
The interaction invariants are computed by the method explained in
\cite{dfinder}. Interaction invariants are over-approximations of the
global state space allowing us to disregard certain tuples of local
states as unreachable. As an illustration, consider the interactions
invariant for the running example when the controller is interacting
with one worker:
\begin{align*}
  \iim\big(\{(a\mid b_1), (c \mid d_1)\}\big) & = (l_{11} \vee lc_{2})
  \,\wedge\, (l_{21}\vee lc_{0}\vee lc_{1}).
\end{align*}
The invariant is given in conjunctive normal form to stick to the
formalism in \cite{dfinder, dfinderJ}. Every disjunction corresponds
to the so called notion of ``initially marked traps'' in an underlying
Petri net associated to our model. Intuitively, a trap in Petri nets
is a set of places which always contains tokens if they have tokens
initially.

We note that the proposed\footnote{The rule \vr\; is generic enough to
  work with other types of invariants. For example, one could use any
  over-approximation of the reachable set in the case of component
  invariants, however, this comes at the price of losing precision.}
component and interaction invariants are inductive invariants.  A
state predicate is called {\em inductive} for a component or system
$B$ if, whenever it holds for a state $s$ of $B$ it equally holds for
any of its successors $s'$.  That is, the validity of an inductive
predicate is preserved by executing any transition, timed or discrete.
An inductive predicate which moreover holds at initial states is an
(inductive) invariant.  Trivially, such a predicate holds in all
reachable states.

As for \textbf{component properties}, we are interested in arbitrary
invariant state properties that can be expressed as boolean
combinations of ``$\mathit{at}(l)$'' predicates and clock constraints.
Invariant properties include generic properties such as mutual
exclusion, absence of deadlock, unreachability of ``bad'' states, etc.
As a simple illustration consider the property $lc_1 \rightarrow
\bigvee_i (y_i - x \geq 4n - 4)$, discussed for our running example
introduced in Section~\ref{sec:intro}.  As a more sophisticated
example, consider \textit{absence of deadlock}.  Intuitively, a timed
system with a set of interactions $\gamma$ is \textit{deadlocked} when
no interaction in $\gamma$ is enabled.  Absence of deadlock is
therefore expressed as the disjunction $\vee_{\alpha \in \gamma}
enabled(\alpha)$. As for the enabledness predicate, we borrow it from
\cite{tripakis99:progress} where it is essentially constructed from
the syntactic definition of the timed system. More precisely, for an
interaction $\alpha$, $\en(\alpha)$ is $\vee_t \en(t)$, with $t$ being
a transition triggered by $\alpha$.  In turn, for $t = \big(\bar{l},
(\alpha, g, r), \bar{l}'\big)$, $\en(t)$ is defined using elementary
operations on zones as $\bar{l} \wedge \swarrow (g \cap
[r]\inv(\bar{l}') \cap \inv(\bar{l}))$, where $\swarrow \zeta$ is the
backward diagonal projection of $\zeta$, $[r]\zeta$ is the set of
valuations $\mathbf{v}$ such that $\mathbf{v}[r]$ is in $\zeta$.



\section{Timed Invariant Generation}
\label{s:bm}

As explained in the introduction, a direct application of the
compositional verification rule (VR) may not be useful in itself in
the sense that the component and the interaction invariants alone are
usually not enough to prove global properties, especially when such
properties involve relations between clocks in different
components. More precisely, though component invariants encode timings
of local clocks, there is no direct way -- the interaction invariant
is orthogonal to timing aspects -- to constrain the bounds on the
differences between clocks in different components. To give a concrete
illustration, consider the property $\varphi_{\mathit Safe} = (lc_1
\wedge l_{11} \rightarrow x \leq y_1)$ that holds in the running
example with one worker. We note that if this property is satisfied,
it is guaranteed that the global system is not deadlocked when the
controller is at location $lc_1$ and the worker is at location
$l_{11}$.  It is not difficult to see that $\varphi_{\mathit Safe}$
cannot be deduced from $\ic(\ctn) \wedge \ic(\cwkO) \wedge
\iim\big(\{(a\mid b_1), (c \mid d_1)\}\big)$ as no relation can be
established between $x$ and $y_1$.

\subsection{History Clocks for Actions}

In this section, we show how we can, by means of some auxiliary
constructions, apply (VR) more successfully. To this end, we ``equip''
components (and later, interactions) with \textit{history clocks}, a
clock per action; then, at interaction time, the clocks corresponding
to the actions participating in the interaction are reset. This basic
transformation allows us to automatically compute a new invariant of
the system with history clocks. This new invariant, together with the
component and interaction invariants, is shown to be, after projection
of history clocks, an invariant of the initial system.

\begin{defi}[Components with History Clocks] Given component $\cn = (L, A,
  \X$, $T$, $\inv$, $s_0)$, its extension with history clocks is the component $
  \cn^h = (L, A, \X \cup \hp, T^h, \inv, s_0^h)$ where
\begin{itemize}
\item $\hp = \{\te\} \cup \{h_a \mid a \in A\}$ is the set of history
  clocks,
\item $T^h = \big\{\big(l, (a, g, r \cup \{h_a \}), l'\big) \mid \big(l,
  (a, g, r), l'\big) \in T\big\}$,
\item $s_0^h = (l_0,c_0^h)$, where $c_0^h = (c_0 \wedge \te = 0 \wedge
  \bigwedge_{a \in A} h_a > 0)$, given $s_0 = (l_0, c_0)$.
\end{itemize}
\label{def:cha}
\end{defi}

The clock $\te$ measures the time from the initialisation.  This clock
equals 0 in $s_0^h$ and is never tested or reset.  Due to this very
restricted use, the same clock $\te$ can be consistently used (shared) by
all components $B^h$ and consequently, allows to capture clock constraints
derived from the common system initialisation time.

Every history clock $h_a$ measures the time passed from the last
occurrence of action $a$.  These history clocks are initially strictly
greater than $0$ and are reset when the corresponding action is
executed. As a side effect, whenever $h_a$ is strictly bigger than
$\te$, we can infer that the action $a$ has not been (yet) executed.
This initialisation scheme allows a more refined analysis precisely
because we can distinguish between actions which were executed and
those which were not.

\mycomment{ In fact,
  if the history clocks were initialised to be equal to $0$ at the
  start time, the global invariant would allow to reflect that all the
  actions have occurred at instant $0$, which is generally
  spurious. In addition, in the following subsections, the method
  proposes an invariant expressing some \textit{separations} between
  the history clocks in some cases. This would induce inaccuracy if
  the history clocks are supposed all equal initially.  Following this
  reasoning, we initialised the history clocks to be strictly positive
  at start time, without giving them concrete values.  }

Since there is no timing constraint involving history clocks, these have no
influence on the behaviour. The extended model is, in fact, bisimilar to
the original model. Moreover, any invariant of the extended model of
$\cn^h$ corresponds to an invariant of original component.  By abuse of
notation, given set of actions $A = \{a_1,...,a_m\}$ use $\exists \hp$ to
stand for $\exists h_{a_1} \exists h_{a_2} \dots \exists h_{a_m} \exists
\te$.

\begin{prop} \label{p:tah} \hfill
  \begin{enumerate}
  \item If $\Phi^h$ is an invariant of $B^h$ then $\Phi = \exists \hp.
    \Phi^h$ is an invariant of $B$.
  \item If $\Phi^h$ is an invariant of $B^h$ and $\Psi^h$ an inductive
    assertion of $B^h$ expressed on history clocks $\hp \setminus \{ h_0
    \}$ then $\Phi = \exists \hp. (\Phi^h \wedge \Psi^h)$ is an invariant of $B$.
  \end{enumerate}
\end{prop}

\proof (1) It suffices to notice that any symbolic state $(l,
\zeta^h)$ in the reachable set $\reach_{\cn^h}(s_0^h)$ corresponds to
a symbolic state $(l, \zeta)$ in the reachable set $\reach_{\cn}(s_0)$
such that $\zeta$ is the projection of $\zeta^h$ to clocks in $\X$,
that is $\zeta \equiv \exists \hp. \zeta^h$.  Henceforth, $\exists
\hp.  \reach_{\cn^h}(s_0^h) \equiv \reach_{\cn}(s_0)$.  Moreover, for
any invariant $\Phi^h$ of $\cn^h$ it holds $\exists
\hp. \reach_{\cn^h}(s_0^h) \subseteq \exists \hp. \Phi^h$.  By
combining the two facts, we obtain that $\Phi$ is an invariant of
$\cn$. \\
\hspace*{1cm} (2) Consider the modified component with history clocks
$\cn^h_\Psi$ defined as $B^h$ but with initial configuration $(l_0,
c_0^h \wedge \Psi^h)$.  This initial configuration is valid, as
$\Psi^h$ constrain exclusively clocks in $\hp$ whereas $c_0^h$ leaves
all of them unconstrained.  Now, it can be easily shown that $\Phi^h
\wedge \Psi^h$ is an invariant of $\cn^h_\Psi$.  Then, following the
same reasoning as for point (1) we obtain that $\exists \hp.  (\Phi^h
\wedge \Psi^h)$ is an invariant of $\cn$.  \qed

The only operation acting on history clocks is reset. Its effect is
that immediately after an interaction takes place, all history clocks
involved in the interaction are equal to zero. All the remaining ones
preserve their previous values, thus they are greater than or equal to
those being reset. This basic observation is exploited in the
following definition, which builds, recursively, all the inequalities
that could hold given an interaction set $\gamma$.

\begin{defi}[Interaction Inequalities for History Clocks]
\label{def:eqs}
Given an interaction set $\gamma$, we define the following
interaction inequalities $\eqs(\gamma)$:
\begin{align*}
  \eqs(\gamma) & = \displaystyle{\bigvee_{\alpha \in \gamma}}
  \Big(\displaystyle{\big(\bigwedge_{\substack{a_i, a_j \in \alpha \\ a_k
        \in \actions(\gamma \ominus \alpha)}}} h_{a_i} = h_{a_j} \leq h_{a_k}
  \big) \wedge \eqs(\gamma\ominus\alpha) \Big).
\end{align*}
where $\gamma\ominus\alpha = \{\beta \setminus \alpha \;|\; \beta \in
\gamma \wedge \beta \not\subseteq \alpha \}$ and $\eqs(\emptyset)=\true$.
\end{defi}

The mechanism of history clocks is as follows. When an interaction
$\alpha$ takes place, the history clocks $h_a$ associated to any
action $a \in \alpha$ are reset. Thus they are all equal and smaller
than any other clocks and measure the time passed from the last
occurrence of $a$.

The operation $\gamma \ominus \alpha$ eliminates in any interaction
$\beta$ the actions from $\alpha$. As an illustration, for $\beta =
(a \mid a_1 \mid a_2)$, $\alpha = (a_1 \mid a_2)$, $\gamma = \{\alpha,
\beta\}$, $\gamma \ominus \alpha = \{a\}$. 


We can use the interpreted function ``$\min$'' as syntactic sugar to have a
slightly more compact expression for $\eqs(\gamma)$ as follows:
\begin{align*}
  \eqs(\gamma) & = \displaystyle{\bigvee_{\alpha \in \gamma}}
  \Big(\displaystyle{\bigwedge_{a_i,a_j \in \alpha}} h_{a_i} = h_{a_j} \leq
  \displaystyle{\min\limits_{a_k \in \actions(\gamma \ominus \alpha)}}h_{a_k} \wedge
  \eqs(\gamma\ominus\alpha)\Big). 
\end{align*} 
As an example, for $\gamma = \{(a\mid b_1), (c\mid d_1)\}$ corresponding to
the interactions between the controller and one worker in
Figure~\ref{fig:ncw}, the compact form is:
\begin{align*}
& \big(h_{a} = h_{b_1} \leq \min(h_c, h_{d_1}) \wedge h_c = h_{d_1}\big) \vee \big(h_{c}=h_{d_1}\leq \min(h_a, h_{b_1}) \wedge h_a = h_{b_1} \big).
\end{align*}

\noindent $\eqs(\gamma)$ characterises the relations between history clocks during
any possible execution. It can be shown that this characterisation is, in
fact, an inductive predicate of the extended system with history clocks.

\begin{prop}
\label{p:eqsI}
$\eqs(\gamma)$ is an inductive predicate of $\|_{\gamma} \cn^h_i$.
\end{prop}
\proof Assume $\eqs(\gamma)$ holds in some arbitrary state $s$ of
$\|_{\gamma} \cn^h_i$. We have two categories of successor states for
$s$, namely time successors and discrete successors. Obviously
$\eqs(\gamma)$ holds for all time successors $s'$, as all clocks
progress uniformly and henceforth all the relations between them are
preserved.  Let now $s'$ be a discrete successor of $s$ by an
arbitrary interaction $\alpha$.  As all the history clocks for actions
in $\alpha$ have just been reset, $s'$ satisfies
\begin{align}
\bigwedge_{\substack{a_i, a_j \in \alpha \\ a_k
    \in \actions(\gamma \ominus \alpha)}} 0 = h_{a_i} = h_{a_j} \leq
h_{a_k} \label{eq:si1}
\end{align} To conclude the proof, we need to show that moreover, for
the remaining clocks of actions in $\actions(\gamma \ominus \alpha)$,
they satisfy $\eqs(\gamma \ominus \alpha)$ in $s'$.  Actually, we can
show the additional fact that for any set of interactions $\gamma$ and
for any interaction $\alpha$ the implication $\eqs(\gamma) \rightarrow
\eqs(\gamma \ominus \alpha)$ is valid in any reachable state. This
fact can be simply proven by induction on the size of the set
interactions $\gamma$ following the definition of
$\eqs$. Consequently, assuming that $\eqs(\gamma)$ holds at $s$, it
follows that $\eqs(\gamma\ominus\alpha)$ holds at $s$. Then
$\eqs(\gamma\ominus\alpha)$ also holds at $s'$ because $\alpha$ does
not modify any clock involved in $\gamma \ominus \alpha$ and this
concludes the proof.  \qed

By using Proposition~\ref{p:eqsI} and Proposition~\ref{p:tah}, we can
safely combine the component and interaction invariants of the system with
history clocks with the interaction inequalities.  We can eliminate the
history clocks from $\bigwedge_i \ic(\cn_i^h) \wedge \iim(\gamma) \wedge
\eqs(\gamma)$ and obtain an invariant of the original system. This
invariant is usually stronger than $\bigwedge_i \ic(\cn_i) \wedge
\iim(\gamma)$ and yields more successful applications of the rule (VR).

\begin{cor}
  $\Phi = \exists \hp. (\bigwedge_i \ic(\cn_i^h) \wedge \iim(\gamma) \wedge
\eqs(\gamma))$ is an invariant of $\|_{\gamma} \cn_i$.
\end{cor}

\begin{exa} 
\label{eg:hc}
We reconsider the model of a controller and a worker from
Figure~\ref{fig:ncw}. We show how the generated invariants are enough
to prove the safety property $\varphi_\mathit{Safe} = (lc_1 \wedge
l_{11} \rightarrow x \leq y_1)$ from Section~\ref{sec:intro}. The
invariants for the components with history clocks are computed
precisely as illustrated in Section~\ref{sec:intro}, that is, they
represent zone graphs:
\begin{align*}
 \ic(\ctn^{h}) = 
& (lc_0 \wedge x = \te < h_a \wedge \te < h_c    )\; {\vee}  \\
& (lc_1 \wedge x \leq \te - 4 \wedge  x \leq 4 \wedge \te  < h_a  \wedge \te < h_c  ) \; {\vee}\\
& (lc_1 \wedge x \leq 4 \wedge  x = h_c \leq h_a \leq \te - 8)\; {\vee}\\
& (lc_2 \wedge x \leq \te - 8 \wedge h_a=x \wedge  \te < h_c  ) \; \vee \\
& (lc_2 \wedge x=h_{a} \wedge  h_{c}=h_{a} + 4 \leq \te - 8)
\end{align*}
\begin{align*}
\ic(   \cwkO^{h}) = 
& (l_{11} \wedge y_1 = \te < h_{d_1}  \wedge \te < h_{b_1}  ) \; {\vee}  \\
& (l_{11} \wedge  y_1 = h_{d_1}  \leq h_{b_1} \leq \te - 4) \; {\vee} \\
& (l_{21} \wedge h_{b_1}+4 \leq y_1 =\te    < h_{d_1}   )) \; {\vee} \\
& (l_{21} \wedge y_1 =h_{d_1}  \leq \te - 4 \wedge h_{b_1} \leq h_{d_1} - 4 )
\end{align*}
By using the interaction invariant described in Section~\ref{sec:frtbip}
 and the inequality constraints $\eqs( (a\mid b_1), (c\mid d_1) )$,
 after the elimination of the existential quantifiers in 
\begin{align*}
\big(\exists h_a.\exists h_{b_1}.\exists h_c.\exists
h_{d_1}.\exists h_{0}\big) \ic(\ctn^{h}) \wedge \ic(\cwkO^{h}) \wedge \iim(\gamma)
\wedge \eqs(\gamma)\big)
\end{align*}
we obtain the following invariant $\Phi$\,:
\begin{align*}
\Phi =  
& (l_{11}\wedge lc_{0} \wedge\, \bm{x=y_1}) \vee \\
& \big(l_{11}\wedge lc_{1} \wedge (\bm{y_1=x\,\vee\,x+4 \leq y_1 })\big) \vee\\
& \big(l_{21}\wedge lc_{2} \wedge \bm{(y_1=x+4\,\vee\, x+8 \leq y_1 )}\big).
\end{align*}

We used bold fonts in $\Phi$ to highlight relations between $x$ and
$y_1$ which are not in $\ic(\ctn) \wedge \ic(\cwkO) \wedge
\iim(\gamma)$. It can be easily checked now that $\Phi \rightarrow
\varphi_{Safe}$ holds and consequently, this proves that
$\varphi_{Safe}$ holds for the system.
\end{exa}
\noindent To sum up, the basic steps of our invariant generation method described so far are: 
\begin{enumerate}
\item compute the interaction invariant $\iim(\gamma)$;
\item extend the components $\cn_i$ to components with history clocks
  $\cn_i^h$;
\item compute component invariants $\ic(\cn_i^h)$;
\item compute inequality constraints $\eqs(\gamma)$ for interactions
  $\gamma$;
\item finally, eliminate the history clocks in $\bigwedge_i \ic(
  \cn_i^{h}) \wedge \iim(\gamma) \wedge \eqs(\gamma)$.
\end{enumerate}
We note that, due to the combination of recursion and disjunction,
$\eqs(\gamma)$ can be large. Much more compact formulae can be
obtained by exploiting non-conflicting interactions, i.e.,
interactions that do not share actions.
\begin{prop}
 \label{p:eqsDisj}
 If $\gamma = \gamma_1 \cup \gamma_2$ such that $\actions(\gamma_1) \cap
 \actions(\gamma_2) = \emptyset$ then $\eqs(\gamma) \equiv \eqs(\gamma_1)
 \wedge \eqs(\gamma_2)$.
\end{prop}
\proof By induction on the number of interactions in $\gamma$. In the
base case, $\gamma$ has a single interaction and the property trivially
holds.  For the induction step, for the ease of reading, we introduce
$eq(\alpha)$ and $leq(\alpha,\gamma)$ to denote respectively
$\bigwedge_{a_i,a_j\in \alpha} h_{a_i} = h_{a_j}$ and
$\bigwedge_{\substack{a_i\in \alpha \\ a_k \in \actions(\gamma\ominus
    \alpha)}} h_{a_i} \le h_{a_k}$. $\eqs(\gamma)$ can be rewritten as follows:
\begin{align*}
\eqs(\gamma) 
 & = \displaystyle{\bigvee_{\alpha \in \gamma_1}}eq(\alpha) \wedge leq(\alpha,\gamma) \wedge \eqs((\gamma_1 \cup \gamma_2) \ominus \alpha) \vee 
     \displaystyle{\bigvee_{\alpha \in \gamma_2}}eq(\alpha) \wedge leq(\alpha,\gamma) \wedge \eqs((\gamma_1 \cup \gamma_2) \ominus \alpha) \\
 & \qquad \big(\textup{using }  \gamma_2 \ominus \alpha = \gamma_2 \textup{ for } \alpha \in \gamma_1 \textup{ and by ind. for } \gamma' = (\gamma_1 \ominus \alpha) \cup \gamma_2 \big)\\
 & \equiv \displaystyle{\bigvee_{\alpha \in \gamma_1}}eq(\alpha) \wedge leq(\alpha,\gamma) \wedge \eqs(\gamma_1 \ominus \alpha) \wedge \eqs(\gamma_2) \vee 
     \displaystyle{\bigvee_{\alpha \in \gamma_2}}eq(\alpha) \wedge leq(\alpha,\gamma) \wedge \eqs(\gamma_1) \wedge \eqs(\gamma_2 \ominus \alpha) \\
 & \qquad \big(\textup{using } \bigvee_{\alpha \in \gamma_i} eq(\alpha) \wedge leq(\alpha, \gamma_i) \wedge \eqs(\gamma_i \ominus \alpha) = \eqs(\gamma_i) \textup{ for } i \in \{1, 2\} \big) \\
 & \equiv \eqs(\gamma_1) \wedge \eqs(\gamma_2) \wedge \big(\mor{\alpha \in \gamma_1}leq(\alpha, \gamma_2) \vee \mor{\alpha \in \gamma_2}leq(\alpha, \gamma_1)) \qquad \\
 & \qquad (\textup{using totality of "$\leq$" and disjointness of $\gamma_i$})\\
 & \equiv \eqs(\gamma_1) \wedge \eqs(\gamma_2) 
\end{align*}
\vspace*{-.7cm}
\qed
The following corollary is an immediate consequence of Proposition \ref{p:eqsDisj}.
\begin{cor}
 \label{cor:eqsDisj}
If the interaction model $\gamma$ has only
  disjoint interactions, i.e., for any $\alpha_1, \alpha_2 \in
  \gamma$, $\alpha_1 \cap \alpha_2 = \emptyset$, then
  $\eqs(\gamma) \equiv \displaystyle{\bigwedge_{\alpha \in
      \gamma}}\Big(\displaystyle{\bigwedge_{a_i,a_j \in \alpha}} h_{a_i} =h_{a_j}\Big)$.
\end{cor}
The two interactions in $\gamma = \{(a\mid b_1), (c\mid d_1)\}$ are
disjoint. Thus, we can simplify the expression of $\eqs(\gamma)$ to $(h_{a}
= h_{b_1}) \wedge (h_{c}=h_{d_1})$.


\subsection{History Clocks for Interactions}
\label{sec:methodEx}

The equality constraints on history clocks allow to relate the local
constraints obtained individually on components.  In the case of
non-conflicting interactions, the relation is rather ``tight'', that is,
expressed as conjunction of equalities on history clocks.  In
contrast, the presence of conflicts lead to a significantly weaker
form.  Intuitively, every action in conflict can be potentially used
in different interactions.  The uncertainty on its exact use leads to
a disjunctive expression as well as to more restricted equalities and
inequalities amongst history clocks.

Nonetheless, the presence of conflicts themselves can be additionally
exploited for the generation of new invariants.  That is, in contrast
to equality constraints obtained from interactions, the presence of
conflicting actions enforce disequalities (or separation) constraints
between all interactions using them. In what follows, we show a
generic way of automatically computing such invariants enforcing
differences between the timings of the interactions themselves. To
effectively implement this, we proceed in a similar manner as in the
previous section: we again make use of history clocks and
corresponding resets but this time we associate them to interactions,
at the system level. 
\begin{defi}[System with Interaction History Clocks]
\label{def:sc}
Given a timed system $\|_{\gamma}\cn_i$, its extension with
history clocks for interactions is the timed system $\sysgammah$ where:
\begin{itemize}
\item $\compgamma$ is an auxiliary component $(\{\dl\}, A_\gamma, \ha, T,
  (\dl \mapsto \true), (\dl,\true))$ where:
\begin{itemize}
\item the set of actions $A_{\gamma} = \{a_{\alpha} \mid \alpha \in \gamma\}$
\item the set of interaction history clocks $\ha = \{h_{\alpha} \mid \alpha \in \gamma\}$
\item the set of transitions $T = \{(\dl,( a_{\alpha}, \true, \{ h_{\alpha} \}), \dl)  \mid \alpha \in \gamma\}$
\end{itemize}
\item $\gamma^h = \{ (a_{\alpha} \mid \alpha) \mid \alpha \in \gamma
  \}$ with $(a_{\alpha} \mid \alpha)$ denoting $\{a_{\alpha}\} \cup
  \{a \mid a \in \alpha\}$.
\end{itemize}
\end{defi}

As before, it can be shown that any invariant of $\sysgammah$ corresponds
to an invariant of $\|_{\gamma}\cn_i$.  The history clocks for interactions
do not impact the behaviour and henceforth the two systems are bisimilar.

\begin{prop} \label{p:dtah} \hfill
  \begin{enumerate}
  \item If $\Phi^h$ is an invariant of $\sysgammah$, then $\Phi = \exists \hp \exists \ha. \Phi^h$ is an invariant of $\|_{\gamma}\cn_i$.
  \item If $\Phi^h$ is an invariant of $\sysgammah$ and $\Psi^h$ an inductive predicate of $\sysgammah$
 expressed on history clocks for actions and interactions $\ha \cup \hp \setminus \{\te\}$ then $\Phi = \exists \hp \exists \ha. (\Phi^h \wedge \Psi^h)$ is an invariant of $\|_{\gamma}\cn_i$.
 \end{enumerate}
\end{prop}
\proof Similar to Proposition~\ref{p:tah}.
\qed

We use history clocks for interactions to express additional constraints on
their timing. The starting point is the observation that when two
conflicting interactions compete for the same action $a$, no matter which
one is first, the latter must wait until the component which owns $a$ is
again able to execute $a$. This is referred to as a ``separation
constraint'' for conflicting interactions.

\begin{defi}[Separation Constraints for Interaction Clocks]
\label{def:sep}
Given an interaction set $\gamma$, the induced separation constraints,
$\sep(\gamma)$, are defined as follows:
\begin{align*}
 \sep(\gamma) = \displaystyle{\bigwedge_{a \in Act(\gamma)}}\;\displaystyle{\bigwedge_{\substack{\alpha\neq\beta \in \gamma\\a \in \alpha \cap \beta}}} \mid h_{\alpha} - h_{\beta} \mid \geq k_a
\end{align*} 
where $\mid x \mid$ denotes the absolute value of $x$ and $k_a$ is a
constant computed locally on the component executing $a$, and representing
the minimum elapsed time between two consecutive executions of $a$.
\end{defi}

In our running example the only conflicting actions are $a$ and $c$ within
the controller, and both $k_a$ and $k_c$ are equal to 4. The
expression of the separation constraints reduces to:
\begin{align*}
  \sep( (a\mid b_i)_i, (c\mid d_i)_i) \equiv\; & 
  \displaystyle{\bigwedge_{i\neq j}}|h_{c|d_{i}}-h_{c|d_{j}}|\geq 4\,\wedge
  \displaystyle{\bigwedge_{i\neq j}}|h_{a|b_{i}}-h_{a|b_{j}}|\geq 4.
\end{align*}

\begin{prop} \label{p:cI} 
 Let $$\sep^*(\gamma) = \displaystyle{\bigwedge_{a \in
     Act(\gamma)}}\;\displaystyle{\bigwedge_{\substack{\alpha\neq\beta \in
       \gamma\\a \in \alpha \cap \beta}}} (h_a \le h_{\alpha} \le h_{\beta}
 - k_a \vee h_a \le h_{\beta} \le h_{\alpha} - k_a)$$
We have that:
\begin{enumerate}
\item $\sep^*(\gamma)$ is an inductive predicate of $\sysgammah$.
\item The equivalence $\sep(\gamma) \equiv \exists \hp. \sep^*(\gamma)$ is a valid formula.
\end{enumerate}
\end{prop}
\proof (1) Let us fix an arbitrary term $S(a,\alpha,\beta)$ defined as 
$$S(a,\alpha,\beta) = (h_a \le h_{\alpha} \le h_{\beta}
- k_a \vee h_a \le h_{\beta} \le h_{\alpha} - k_a)$$ Assume
$S(a,\alpha,\beta)$ holds in an arbitrary state $s$ of $\sysgammah$.
Then, it obviously holds for any time successors as well as for any
discrete successors by interactions not containing the action $a$.
For an interaction involving $a$, but different than $\alpha$ and
$\beta$, $h_a$ is reset to zero whereas $h_\alpha$ and $h_\beta$ are
unchanged.  Henceforth, $S(a,\alpha,\beta)$ remains valid as only
$h_a$ changes to 0.  Let consider the situation $\alpha$ is executed
(the case of $\beta$ is perfectly dual).  In this case, both $h_a$ and
$h_\alpha$ are reset to 0, whereas $h_\beta$ is unchanged.  Two
situations can happen:
\begin{enumerate}[label=\({\alph*}]
\item $h_a \le h_{\alpha} \le h_{\beta} - k_a$ holds in $s$.  Then,
  obviously, the same holds in $s'$ where $h_a$ and $h_\alpha$ are reset.
\item $h_a \le h_{\beta} \le h_{\alpha} - k_a$ holds in $s$.  This is the
  interesting case where we need the assumption about the separation time
  $k_a$.  As consecutive executions of $a$ are separated by $k_a$, to
  execute $\alpha$ it must actually hold that $h_a \ge k_a$ in $s$.
  Consequently, $h_\beta \ge k_a$ in $s$, as well as in $s'$ (because
  $h_\beta$ does not change from $s$ to $s'$).  Then, knowing that $h_a =
  h_\alpha = 0$ in $s'$ we have that $h_a \le h_\alpha \le h_\beta - k_a$
  in $s'$.
\end{enumerate}
(2) We can equivalently write 
\begin{align*}
\sep^*(\gamma) 
& \equiv \displaystyle{\bigwedge_{a \in
     Act(\gamma)}}\;\displaystyle{\bigwedge_{\substack{\alpha\neq\beta \in
       \gamma\\a \in \alpha \cap \beta}}} (h_a \le h_{\alpha} \wedge h_a \le h_\beta \wedge \mid h_\alpha - h_\beta \mid \le k_a) 
\end{align*}
\begin{align*}
& \equiv \sep(\gamma) \wedge \displaystyle{\bigwedge_{a \in
     Act(\gamma)}}\;\displaystyle{\bigwedge_{\substack{\alpha\neq\beta \in
       \gamma\\a \in \alpha \cap \beta}}} (h_a \le h_{\alpha} \wedge h_a \le h_{\beta}) 
\end{align*}
and this concludes our proof.\qed

The predicate $\sep(\gamma)$ is expressed over history clocks for
interactions.  Component invariants $\ic(\cn_i^h)$ are however
expressed using history clocks for actions.  In order to ``glue'' them
together in a meaningful way, we need some tighter connection between
action and interaction history clocks. This aspect is addressed by the
constraints $\eqsc$ defined below.
\begin{defi}[$\eqsc$]
\label{def:eqsC}
Given an interaction set $\gamma$, we define $\eqsc(\gamma)$ as follows:
\begin{align*}
\eqsc(\gamma) = \displaystyle{\bigwedge_{a \in Act(\gamma)}} h_a = \min\limits_{{\alpha \in \gamma, a \in \alpha}} h_\alpha.
\end{align*}
\end{defi}

By a similar argument as the one in Proposition~\ref{p:eqsI}, it can be
shown that $\eqsc(\gamma)$ is an inductive predicate of the extended system
$\sysgammah$.  Moreover, there exists a tight connection between $\eqs$ and
$\eqsc$ as given in Proposition~\ref{p:conneqsC}.

\begin{prop} 
\label{p:conneqsC} \hfill
\begin{enumerate}
\item  $\eqsc(\gamma)$ is an inductive predicate of $\sysgammah$.
\item The equivalence $\exists \ha.\eqsc(\gamma) \equiv \eqs(\gamma)$ is a
  valid formula.
\end{enumerate}
\end{prop}
\proof (1) To see that $\eqsc(\gamma)$ is an inductive predicate it
suffices to note that the predicate is preserved by time progress
transitions and for any discrete action $a$, there is always an interaction
$\alpha$ containing $a$ such that $h_a$ and $h_{\alpha}$ are both reset in
the same time.\\
(2) The proof follows directly from the definitions of $\eqs(\gamma)$ and
$\eqsc(\gamma)$.  Consider that $\gamma= \{\alpha_1, \alpha_2, ..., \alpha_m
\}$.  We have the following equivalences:
\begin{align*}
\exists \ha. \eqsc(\gamma) 
& \equiv \exists \ha. \bigvee\limits_{\alpha_{k_1} \prec \alpha_{k_2} \prec ... \prec \alpha_{k_m}} 
\big( h_{\alpha_{k_1}} \le h_{\alpha_{k_2}} \le ... \le h_{\alpha_{k_m}} \wedge \eqsc(\gamma) \big) \\
& \qquad \textup{ (by choosing an arbitrary ordering $\prec$ on interactions) } \\
& \equiv  \exists \ha. \bigvee\limits_{\alpha_{k_1} \prec \alpha_{k_2} \prec ... \prec \alpha_{k_m}} 
\big( h_{\alpha_{k_1}} \le h_{\alpha_{k_2}} \le ... \le h_{\alpha_{k_m}} \wedge \\
& \qquad \bigwedge_{a \in \alpha_{k_1}} (h_a = h_{\alpha_{k_1}}) \wedge 
 \bigwedge_{a \in \alpha_{k_2} \setminus \alpha_{k_1}} (h_a = h_{\alpha_{k_2}}) \wedge ... 
 \bigwedge_{a \in \alpha_{k_m} \setminus \alpha_{k_1} ... \alpha_{k_{m-1}}} (h_a = h_{\alpha_{k_m}}) \big) \\
& \qquad \textup{ (by expanding the definition of $\eqsc(\gamma)$ along the chosen order) }
\end{align*}
\begin{align*}
& \equiv \exists \ha. \bigvee\limits_{\alpha_{k_1} \prec \alpha_{k_2} \prec ... \prec \alpha_{k_m}} 
\big( h_{\alpha_{k_1}} \le h_{\alpha_{k_2}} \le ... \le h_{\alpha_{k_m}} \wedge 
\bigwedge_{\ell = 1}^m \bigwedge_{a \in \alpha_{k_\ell}\setminus \alpha_{k_1} ... \alpha_{k_{\ell-1}}} (h_a = h_{\alpha_{k_\ell}}) \big) \\
& \qquad \textup{ (by rewriting to a more compact form) } \\
& \equiv \bigvee\limits_{\alpha_{k_1} \prec \alpha_{k_2} \prec ... \prec \alpha_{k_m}} \exists \ha. 
\big( h_{\alpha_{k_1}} \le h_{\alpha_{k_2}} \le ... \le h_{\alpha_{k_m}} \wedge 
\bigwedge_{\ell = 1}^m \bigwedge_{a \in \alpha_{k_\ell}\setminus \alpha_{k_1} ... \alpha_{k_{\ell-1}}} (h_a = h_{\alpha_{k_\ell}}) \big) \\
& \qquad \textup{ (by distributing the existential quantifiers over the disjunction) } 
\end{align*}
\begin{align*}
&  \equiv \bigvee\limits_{\alpha_{k_1} \prec \alpha_{k_2} \prec ... \prec \alpha_{k_m}} 
\bigwedge_{\ell = 1}^m \bigwedge_{\substack{a_i,a_j \in \alpha_{k_\ell} \setminus \alpha_{k_1} ... \alpha_{k_{\ell-1}} \\ 
  a_k \not\in \alpha_{k_1} ... \alpha_{k_\ell}}} (h_{a_i} = h_{a_j} \le h_{a_k}) \equiv \eqs(\gamma) 
\end{align*}
$\qquad \qquad \qquad\qquad$(by eliminating the existential quantifiers) \qed\medskip

\noindent From Propositions~\ref{p:conneqsC},~\ref{p:dtah}, and~\ref{p:cI}, it
follows that $\exists \hp \exists \ha. (\bigwedge_i \ic(\cn_i^h) \wedge
\iim(\gamma) \wedge \eqsc(\gamma) \wedge \sep(\gamma))$ is an invariant of
$ \|_{\gamma}\cn_i$. This new invariant is in general stronger than
$\exists \hp. (\bigwedge_i \ic(\cn_i^h) \wedge \iim( \gamma) \wedge
\eqs(\gamma))$ and it provides better state space approximations for timed
systems with conflicting interactions.

\begin{cor}
$\Phi = \exists \hp \exists \ha. (\bigwedge_i \ic(\cn_i^h) \wedge
\iim(\gamma) \wedge \eqsc(\gamma) \wedge \sep(\gamma))$ is an invariant of  $\|_{\gamma}\cn_i$.
\end{cor}

\begin{exa}
  To get some intuition about the invariant generated using separation
  constraints, let us reconsider the running example with two
  workers. The subformula which we emphasise here is the conjunction
  of $\mathcal{E}^{*}$ and $\mathtt{\mathcal{S}}$. The interaction invariant is:
 \begin{align*}
\iim(\gamma) = 
& (l_{11}\vee lc_{1}\vee lc_{2}) \wedge (l_{12}\vee lc_{1}\vee lc_{2})  \wedge 
 (lc_{2}\vee l_{11}\vee l_{12})\wedge(lc_{0}\vee lc_{1}\vee l_{21}\vee l_{22})
\end{align*}
The components invariants are:
\begin{align*}
\ic(\ctn^{h}) = 
& (lc_0 \wedge x = \te \wedge \te < h_a \wedge \te < h_c )\; {\vee}  \\
& (lc_1 \wedge x \leq \te - 8 \wedge  x \leq 4 \wedge \te < h_a  \wedge \te < h_c ) \; {\vee} \\
& (lc_1 \wedge x \leq 4 \wedge  x = h_c \leq h_a \leq \te - 12)\; {\vee} \\
& (lc_2 \wedge x \leq \te - 12 \wedge h_a=x \wedge \te < h_c ))\; {\vee} \\
& (lc_2 \wedge x=h_{a} \wedge  h_{c}=h_{a} + 4 \leq \te - 12)\\
\end{align*}
\begin{align*}
 \ic(\cwk_i^{h}) = 
& (l_{1i} \wedge y_i =\te  \wedge \te < h_{d_i} \wedge \te < h_{b_i}    ) \; {\vee}  \\
& (l_{1i} \wedge  y_i = h_{d_i}  \leq h_{b_i} \leq \te - 8) \; {\vee} \\
& (l_{2i} \wedge y_i \geq h_{b_i}+8 \leq \te < h_{d_i}  )) \; {\vee} \\
& (l_{2i} \wedge y_i =h_{d_i}  \leq \te - 8 \wedge h_{b_i} \leq h_{d_i} - 8 ) \\
\end{align*}
The inequalities for action and interaction history clocks are:
\begin{align*}
\eqsc(\gamma) =
& (h_{b_1}=h_{a|b_{1}}) \wedge (h_{b_2}=h_{a|b_{2}}) \wedge\,\, (h_{a}=\min_{i=1,2}(h_{a|b_{i}})) \wedge \\
& (h_{d_1}=h_{c|d_{1}}) \wedge (h_{d_2}=h_{c|d_{2}}) \wedge\,\, (h_{c}=\min_{i=1,2}(h_{c|d_{i}})) 
\end{align*}
By recalling the expression of $\sep(\gamma)$ we obtain that:
\begin{align*}
& \exists\mathcal{H}_{\gamma}.\mathcal{E}^{*}\left(\gamma\right)\wedge\mathcal{S}(\gamma)\,\, = (|h_{b_{2}}-h_{b_{1}}|\geq 4\,\wedge|h_{d_{2}}-h_{d_{1}}|\geq 4)
\end{align*}
and thus, after quantifier elimination in 
\begin{align*}
\exists \hp \exists \hist_{\gamma}.(\ic(\ctn^{h}) \wedge \bigwedge_i \ic(\cwk_i^{h}) \wedge \iim(\gamma) \wedge \eqsc(\gamma) \wedge \sep(\gamma))
\end{align*}
after simplification, we obtain the following invariant $\Phi$:
\begin{align*}
\Phi = 
& \big(l_{11}\wedge l_{12}\wedge lc_{0} \wedge\, {x=y_1=y_2}\big) \vee \\
& \big(l_{11}\wedge l_{12}\wedge lc_{1} \wedge {x \leq 4 \wedge( y_1=y_2 \geq x+8 \vee\, }  \\
& \qquad \qquad \qquad \qquad \qquad \quad {(y_1=x\, \bm{\wedge y_2-y_1 \geq 4}) } \vee \\
& \qquad \qquad \qquad \qquad \qquad \quad {(y_1 \geq x+8 \wedge y_1-y_2 \geq 8)} \vee\\
& \qquad \qquad \qquad \qquad \qquad \quad {(y_2=x\, \wedge \bm{y_1-y_2 \geq 4})} \vee\\
& \qquad \qquad \qquad \qquad \qquad \quad {(y_2 \geq x+8 \wedge y_2-y_1 \geq 8)  }) \big) \vee\\
& \big(l_{21}\wedge l_{12}\wedge lc_{2} \wedge {y_1\geq x+8 \wedge ( (y_2 \geq x+4 \wedge \bm{|y_1-y_2| \geq \,4 } )}\vee\\
& \qquad \qquad \qquad \qquad  \qquad  \qquad         {  y_2 \geq x+12 \, }) \big) \vee\\
& \big(l_{11}\wedge l_{22}\wedge lc_{2} \wedge {y_2\geq x+8 \wedge ( (y_1 \geq x+4 \wedge \bm{|y_1-y_2| \geq \,4 } )}\vee\\
& \qquad \qquad \qquad \qquad  \qquad  \qquad         {  y_1 \geq x+12 \, }) \big)
\end{align*}
We emphasised in the expression of $\Phi$ the newly discovered
constraints. All in all, $\Phi$ is strong enough to prove that the
system is deadlock free.
\end{exa}

We conclude the section with a discussion about the computation of the
separation constants $k_a$.  A simple but incomplete heuristics to
test that a given value $k_a$ is a correct separation constraint for
an action $a$ is as follows.  Consider all paths connecting two
transitions (not necessarily distinct) labelled by $a$.  If on every
such path, there exists a clock $x$ which is reset and then tested in
a guard $x \ge ct$, with $ct \ge k_a$ then, it is safe to conclude
that actually $k_a$ is a correct separation value.  Nonetheless,
alternative methods to exactly compute $k_a$ have been already
proposed in the literature.  For details, the interested reader can
refer, for instance, to \cite{courcoubetis92} which reduces this
computation to finding a shortest path in a weighted graph built from
the zone graph associated to the component.



\section{Improving (VR) - Three Heuristics}
\label{sec:heuristics}

We describe and elaborate on heuristics allowing to strengthen the
generated invariants and to reduce the generation time. These
heuristics have been successfully applied on our case studies
considered later in Section~\ref{sec:impl}.

\subsection{Refining conflicting interactions}
\label{sec:h0inv}

The initialisation of the history clock $\te$ provides a convenient
way to express and reason about invariants relating occurences of
various actions and interactions at execution.  The assertion
$h_\alpha \leq \te$ has the intuitive meaning that ``$\alpha$ has been
executed''.  We describe below a new family of invariants providing a
finer characherisation for the execution of conflicting interactions
and related actions.

We fix $a$ as a potential conflicting action within some component $B$
= $(L, A, T, \X, \inv)$. We define the set of preceding actions
$Prec(a)$ as all actions of $B$ that can immediately precede $a$ in an
execution, formally $Prec(a) = \{a'\in A \mid \exists l, l', l'' \in
L. \, l \transit{a'} l', l'\transit{a} l''\}$. For any two conflicting
interactions $\alpha_1, \alpha_2$ involving $a$, the following
assertion:
\[ h_{\alpha_1} \leq \te \wedge h_{\alpha_2} \leq \te \Rightarrow
\bigvee\limits_{a'\in Prec(a)} h_{a'} \leq \te \] is an
invariant. Intuitively, the assertion states that whenever $\alpha_1$
and $\alpha_2$ have both been executed (implying that $a$ has also
been executed two or more times), at least one of the preceding
actions of $a$ must also has been executed. We remark that the
invariant above is rather weak and can be implied by the component
invariant $CI(B)$ and the glue invariant $\mathcal{E}^{*}$ in many
situations. In fact, whenever $a$ is an action which {\em is not
  enabled at the initial location} of $B$, the component invariant
$CI(B)$ implies that
\[h_a \leq \te \Rightarrow \bigvee\limits_{a'\in Prec(a)} h_{a'} \leq
\te.\] This states that whenever $a$ has been executed, at least one
of its preceding actions has been executed as well. Knowing moreover
that $h_a = \min_{a \in \alpha} h_\alpha$, we can then infer the
invariant above.

Nonetheless, if $a$ is an action that is enabled at the initial
location, the newly proposed invariant is stronger and cannot be
derived as shown before.  In this case, $a$ can be actually executed
once while none of its predecessors has been executed yet.  The
component invariant alone does not relate anymore the execution of $a$
to the execution of its preceding actions.  Moreover, the component
invariant considers always the last occurence of $a$ and has no means
of distinguishing cases where $a$ has been executed only once or more
often.  This information can sometimes be re-discovered when
interaction history clocks $h_{\alpha_1}$, $h_{\alpha_2}$ are taken
into account, henceforth, leading to the proposed invariant. A
concrete illustration is provided later in Section~\ref{sec:impl}.


\subsection{Invariant computation using regular expressions}
\label{sec:regex}

There exist situations where the computation of component invariants
can be extremely costly. In particular, for untimed components
extended with history clocks, their zone graphs will most likely have
an exponential size.  In fact, due to history clocks, the zones will
record the order of (the last) occurences of actions, and there could
be exponentially many of them, reachable at different locations. We
note that, in timed components, clocks \emph{restrict} the dynamics of
the components, consequently, it cannot be the case that \emph{all} the
orders are possible.


The above observation suggests (and was confirmed by our experiments)
that applying the same methodology for computing component invariants
(based on the reachability graph of the corresponding components with
history clocks) regardless of the components being timed or not leads
to large formulae when possibly shorter ones exist.

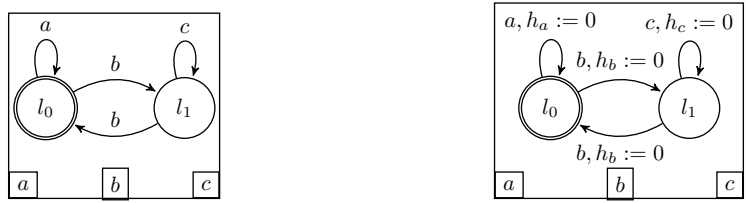
\begin{figure}[htbp]
\input{regexpTikz.tex}
  \caption{\label{fig:regex} An untimed component (left) and its extension
    with history clocks (right).}
\end{figure}

\begin{exa}\label{ex:regex-1} Consider the untimed component
  presented in Figure \ref{fig:regex}
  (left) and its extension with history clocks (right).  The entire
  zone graph reachable from $\langle l_0, \zeta_0\rangle$, with
  $\zeta_0 = (\te=0, h_{a,b,c}>0)$ has $6$ symbolic states.
  Therefore, the component invariant is expressed as a disjunction of
  $16$ terms, $9$ of them are related to location $l_0$ and $7$ are related 
to location $l_1$.
\end{exa}

We recall that untimed automata have elegant and compact encodings as
regular expressions. This basic fact can be exploited in order to
provide an alternative computation method for component invariants.
More concretely, given an untimed component $B=(L, A, T)$ we show how
to automatically compute the invariant describing the relations
between the history clocks of $B^h$ at some location $\ell$, from the
language accepted by $B$ at some designated location $\ell$.  The
first key observation is that only the last occurrence of each action
should be retained. This implies that it is safe to abstract, with
respect to last occurrences, the regular expression characterising the
language accepted at the chosen control location. The second key
observation is that, regular expressions in some restricted form, can
be used to directly generate less constraints on the history
clocks. Our regular
expression based method can be therefore summarised as follows:
\begin{enumerate}
\item construct the regular expression $E_\ell$ representing the language
  accepted by $B$ at location $\ell$,
\item abstract $E_\ell$ with respect to the last occurence retention
  towards some \emph{restricted form} $E_\ell^\sharp = \sum_{i}
  e_i^\sharp$ where, every $e_i^\sharp$ contains each action at most
  once, and does not contain nested *-operators,
\item generate from every $e_i^\sharp$ a characteristic formula on history
  clocks $\phi(e_i^\sharp)$ and obtain as invariant for $B$ the assertion
  $\ell \Rightarrow \vee_i \phi(e_i^\sharp)$.\medskip
\end{enumerate}

\noindent The first step is well known for finite automata and will not be detailed
here. For the second abstraction step, the key ingredients are the
simplification rules in Figure~\ref{fig:rules}.
\begin{figure}[htp]
\begin{align*}
& \textbf{Rule 1 } \textup{[Last Occurrence Retention]:} & E\cdot a \longrightarrow  (E \smallsetminus a) \cdot a \\
& \textbf{Rule 2 } \textup{[Back-unfolding]:} & E^{*} \longrightarrow  (E^{*}\cdot E)+\varepsilon
\end{align*}
\caption{Simplification Rules}
\label{fig:rules}
\end{figure}

Rule 1 eliminates all but the last occurrence of the trailing $a$ symbol 
from a regular expression of the form $E\cdot a$.  The ``$\smallsetminus$''
denotes a syntactic \textit{elimination operator} defined structurally on
expressions as follows. Let $a$ and $x$ be two symbols and $E$, $E_1$ and
$E_2$ be arbitrary regular expressions.
\begin{eqnarray*}
 \epsilon \smallsetminus a &= & \epsilon \\
 x \smallsetminus a &= &\begin{cases}
                      \epsilon \textrm{ if }  x= a \\
                       x \textrm{ if }  x \neq a  \\
                        \end{cases} \\
 ( E_1+E_2 ) \smallsetminus a &= &( E_1 \smallsetminus a)+ (E_2 \smallsetminus a) \\ 
 ( E_1.E_2 ) \smallsetminus a  &=& ( E_1 \smallsetminus a). (E_2 \smallsetminus a) \\
  E^* \smallsetminus a &= & ( E \smallsetminus a)^* \\
\end{eqnarray*}

Rule 2 simply unfolds *-expressions once.  By using this rule and other
basic manipulation of regular expressions, further simplification
opportunities for Rule 1 are enabled.

\begin{exa} Let us consider again the example presented in Figure
  \ref{fig:regex}.  The language accepted at $l_1$ is defined as
  $(a+bc^*b)^*bc^*$.  This expression is progressively abstracted into
  the restricted form as follows:
  \begin{align*}
    (a+bc^*b)^*bc^* & \leadsto (a+c^*)^*bc^*  \qquad & \textup{ (by Rule 1) } \\
    & \equiv (a+c^*)^*b(c^*c+\epsilon) \qquad & \textup{ (by Rule 2) } \\
    & \equiv (a+c^*)^*bc^*c + (a+c^*)^*b \qquad & \textup{ (by splitting the last +) } \\
    & \leadsto (a+\epsilon)^*bc + (a+c^*)^*b & \qquad \textup{ (by Rule 1) } \\
    & \equiv a^*bc + (a+c)^*b \qquad & \textup{ (by standard transformation) }
  \end{align*}
\label{eg:rex}
\end{exa}\medskip

\noindent In the example above, we have applied the iterative strategy
consisting of (1) choosing symbols from right to left and applying
Rule 1 until no longer possible and then (2) applying Rule 2 to unfold
the rightmost *-expression and split the incoming +.  It can be shown
that such a strategy always terminates with expressions in the
restricted form. Intuitively, what happens is that Rule 2 splits
larger expressions into smaller ones and, further, for each of these
Rule 1 eliminates repetitions of symbols.

For the third step, we construct from a regular expression $e^{\sharp}$ in
restricted form an equivalent formula $\phi(e^{\sharp})$ on history
clocks. This formula represents {\em exactly} the set of orders on
actions (the strings) encoded by the regular expression:
\[\phi(e^{\sharp}) \equiv \bigvee\limits_{\substack{a_1...a_n \in L(e^{\sharp})\\\textup{distinct }a_1,\dots,a_n}} \big(\te \geq h_{a_1} \geq ... \geq
h_{a_n} \wedge \bigwedge_{c\not=a_1,...,a_n} h_c > \te\big)\] where
$L(e^{\sharp})$ is the language of $e^{\sharp}$. We note that since we
only consider words with distinct symbols, they are finitely many and
the disjunction is finite as well.

As an illustration, let $e^{\sharp}$ be the regular expression in the
restricted form $a^*bc + (a+c)^*b$ obtained in
Example~\ref{eg:rex}. The finite words on which $\phi(e^{\sharp})$
builds upon are $abc$ and $bc$ (from $a^*bc$) and $acb, cab, cb, ab,
b$ from $(a+c)^*b$. By applying the above encoding, we obtain:
\begin{align*}
& (h_0 \geq h_a \geq h_b \geq h_c)\, \vee\, (h_a > h_0 \geq h_b \geq h_c)\, \vee & (\textup{corr. to }abc, \textup{resp. }bc)\\ 
& (h_0 \geq h_a \geq h_c \geq h_b)\, \vee\, (h_0 \geq h_c \geq h_a \geq h_b)\, \vee & (\textup{corr. to } acb, \textup{resp. }cab)\\
& (h_a > h_0 \geq h_c \geq h_b)\, \vee\, (h_c > h_0 \geq h_a \geq h_b)\, \vee & (\textup{corr. to }cb, \textup{resp. }ab)\\
& (h_0 \geq h_b \wedge h_c, h_a > h_0) & (\textup{corr. to }b)
\end{align*}

\noindent Such encodings are, in fact, invariants. Intuitively, the inequalities
in $\phi(e^{\sharp})$ reflect precisely the order in which the last
action occurences have taken place.

\begin{prop}
  Let $\cn$ be an untimed component, $E_l$ the regular expression
  characterising the language accepted by $\cn$ at location $l$, and
  $E_l^\sharp$ be the result of applying the simplification rules.  We
  have that $\bigvee_l (l \wedge \phi(E_l^{\sharp}))$ is an invariant
  of $\cn^h$.
\end{prop}
\mycomment{de la Marius: (existing text for equivalence $L(e)$ and
  $L(e^#)$ $\dots$ Now, for regular expressions $e^#$ in restricted
  form we can prove the following property.  For every word $w$ in
  $L(e^\sharp)$, the restricted sub-word $w_loc$ obtained from $w$ by
  removing all but last occurrences of every symbol belongs to
  $L^(e^\sharp)$ as well.  Henceforth, one can enumerate over all last
  occurrence words $w_loc$ by simply considering all accepted words of
  $L(e^\sharp)$ having distinct symbols.  $\dots$ (existing text for
  constructing the constraints $\Phi(e^\sharp))$ } 

\proof (sketch) The local component invariant at some location $l$ is
precisely characterised by the orders of the last occurrences of
actions on traces reaching $l$. To show that these orders are captured
by $\phi(E_l^{\sharp})$, it suffices to note that, on the one hand,
$E_l$ and $E_l^{\sharp}$ preserve the language of the last occurrences
of actions. This follows from the simplification rules. As for regular
expressions $e^{\sharp}$ in restricted form we can prove the following
property.  For every word $w$ in $L(e^\sharp)$, the restricted
sub-word $w_{loc}$ obtained from $w$ by removing all but last
occurrences of every symbol belongs to $L(e^\sharp)$ as well.
Henceforth, one can enumerate over all last occurrence words $w_{loc}$
by simply considering all accepted words of $L(e^\sharp)$ having
distinct symbols. To conclude the proof we only need to note that
the inequalities in $\phi(E_l^{\sharp})$ encode the enumeration of all
possible words corresponding to traces of $\cn^h$ ending at $l$. \qed

We can exploit the structure of regular expressions in restricted form
to optimise the technique described above even further. To illustrate
this, we consider the regular expression $(b_1 + ... + b_m)^*
a_1...a_n$ in restricted form (whenever $a_1, ..., a_n,$ $b_1, ...,
b_m$ are distinct).  The corresponding formula on history clocks is
\[\te \geq h_{a_1} \geq ... \geq h_{a_n} \wedge h_{b_1} \ge h_{a_1} \wedge ... \wedge h_{b_m} \ge h_{a_1} 
\wedge \bigwedge\limits_{c\not=a_i,b_j} h_c > \te. \] The first part
encodes the ordering constraints on the {\em mandatory} string
$a_1...a_n$. All these actions occur (consequently, their history
clocks are smaller than $\te$) in this precise order. The second part
considers constraints on occurences of $b_j$ actions, which are {\em
  optional}: if some occur, their executions are unconstrained by each
other, however, they take place before $a_1$.  Finally, the last part
deals with actions $c$ which do not appear in the regular expression.
For all of them, their history clocks should be strictly greater than
$\te$.  We remark that, for this particular example, the obtained
formula has linear size with respect to the size of the regular
expression. In contrast, the number of strings encoded (i.e., whenever
restricted to last occurrences of symbols) is exponential, with
respect to the number of $b$ actions.  The construction above can be
generalised for arbitrary restricted regular expressions without much
difficulty.  The resulting formula remains of polynomial size (at
worse quadratic) with respect to the size of the restricted regular
expression provided as input.

\begin{exa}
  Following the approach described above, the regular expression in
  the restricted form $a^*bc + (a+c)^*b$ translates into:
  \[ (h_0 \geq h_b \geq h_c \wedge h_a \geq h_b) \vee (h_0 \geq h_b
  \wedge h_a \geq h_b \wedge h_c \geq h_b ) \] We note this expression
  is significantly smaller, yet logically equivalent to the
  disjunction of $7$ distinct terms corresponding to symbolic zones
  reached at $l_1$ as initially presented in Example \ref{ex:regex-1}.
\end{exa}

To sum up, we described a heuristic which can be applied to untimed
components to automatically compute an invariant with a reasonable
enough size to be handled by existing SMT solvers. Given an untimed
component $B$, our heuristic makes use of the regular expressions
characterizing the language accepted by $B$ to avoid a direct
construction of the zone graph of $B^h$ which would result in
considerably large invariants.



\subsection{Exploiting Symmetry}
\label{sec:sym}

At a closer examination of the definition of separation constraints in
Section \ref{sec:methodEx}, it can be noticed that it characterises
all possible orderings of conflicting interactions with respect to
permutations.  The size of the corresponding search space is
exponential in the number of conflicting interactions and this, in
turn, may be a bottleneck for the solver. Such situations can and
must be avoided especially in the case of symmetric systems. What we
show next is how the inherent symmetry in the formula can be
eliminated such that the search space becomes considerably smaller.

The use of symmetry has long been addressed, mostly with the intention
of making model-checking more feasible and especially in the context
of parameterised systems
\cite{emersonN95,emersonS96,emersonK00,namjoshi07}.  There the goal is
to show the existence of a small cutoff bound which allows the
reduction of the verification problem from an arbitrary number of
instances to a small, fixed one.  Our context is different, that is,
breaking the symmetry in some of the generated invariants, for an
a priori known number of components.

The types of systems we consider next are formed of a fixed number, be
it $n$, of isomorphic components interacting with a controller, thus
the interactions are binary. Isomorphic components are obtained from a
generic component $\cn$ by attaching an index $i$ (from 1 to $n$) to
all symbols in $\cn$. The resulting component is denoted by
$B_i$. For any $i, j$, $\cn_i$ and $\cn_j$ are
isomorphic\footnote{We note that, by construction, isomorphic
  components cannot have clock constraints involving indices: any
  constraint in a worker $\cn_i$ is obtained from those in $\cn$
  which are oblivious to indices $i$.}. For the ease of reference, we
denote systems like $\ctrl \|^n_{\gamma} \cn_i$ by the letter $M$ and
we use $\exg$ to denote the set of their global executions.

In this framework, the notion of symmetry is intrinsically related to
permutations.  Let $\Pi_n$ denote the group of permutations of $n$.
The application of permutations is defined on the structure of systems
and properties.  For a system $M$ as $\ctrl \|^n_{\gamma} \cn_i$,
and a permutation $\pi$, $\pi(M)$ is defined as $\ctrl
\|^n_{\pi(\gamma)} \pi(\cn_i)$ where $\pi(\cn_i)$ is defined as
$\cn_{\pi(i)}$ and $\pi(\gamma)$ as $\{\pi(\alpha) \mid \alpha \in
\gamma\}$ with $\pi(a_c\mid a_i) = a_c \mid a_{\pi(i)}$ for $\alpha$
an arbitrary binary interaction between an action $a_c$ of $\ctrl$ and
an action $a_i$ of a $B_i$. For an execution $\sigma = \alpha_1, \dots
\alpha_i, \dots \alpha_k$, $\pi(\sigma)$ is defined as $\pi(\alpha_1),
\pi(\alpha_2) \dots \pi(\alpha_i), \dots, \pi(\alpha_k)$. For a global
state $s = (s_c, s_1, \dots, s_n)$, $\pi(s)$ is defined as $(s_c,
s_{\pi(1)}, \dots, s_{\pi(n)})$.  As for system properties $\varphi$, we
restrict to those built (with the usual logical connectors) from clock
constraints and locations, and define:
\[
\pi(\varphi) = 
\begin{cases}
   x_{\pi(i)} \textit{ rop } x_{\pi(j)} & \textup{ if } \varphi = x_i \textit{ rop } x_j \textup{ and } \textit{rop} \in \{<,\leq,=,>, \geq\}\\
  l_{\pi(i)} & \textup{ if } \varphi = l_i \\
  \neg \pi(\varphi_1) & \textup{ if } \varphi = \neg \varphi_1 \\
  \pi(\varphi_1) \textit{ op } \pi(\varphi_2) & \textup{ if } \varphi = \varphi_1 \textit{ op } \varphi_2 \textup{ and } \mathit{op} \in \{\wedge, \vee\}
\end{cases}
\]
where $l_i, x_i$ denote a location, respectively, a clock in $B_i$.

The symmetric systems we consider are symmetric in a ``strong'' sense,
i.e., they are \textit{fully symmetric}. A system $M$ is fully
symmetric if for any $\pi \in \Pi_n$, $\pi(M)$ is syntactically
identical to $M$.  Similarly, a property $\varphi$ is fully symmetric
if for any permutation $\pi$, $\pi(\varphi)$ is equivalent to
$\varphi$.  A property like $l_1 \wedge l_2 \wedge ... \wedge l_n$ is
symmetric. On the contrary, $G = x_1 \leq x_2$
is not as for the permutation $\pi(1) = 2, \pi(2)=1$, $\pi(G) =
x_{\pi(1)} \leq x_{\pi(2)} = x_2 \leq x_1$ which is not equivalent to
$G$.

Symmetric systems have the convenient property that, whenever started
in a symmetric state, for any of its executions $\sigma \in \exg$,
$\pi(\sigma)$ is itself an execution, that is, $\pi(\sigma) \in
\exg$. To see why this is indeed the case, let $\gamma$ be the
interaction set and $\alpha = (a_c\mid a_i)$ an interaction in
$\gamma$. It suffices to note that if $\alpha$ is possible after
$\sigma$, then it is also the case for $\pi(\alpha)$ after
$\pi(\sigma)$.  Note also that, thanks to symmetry, $\pi(\alpha)$ is
in $\gamma$.

The idea behind simplifying the separation constraints $\sep$ is to
break the symmetry by replacing the constraints on absolute values
$\mid h_{\alpha_i} - h_{\alpha_j} \mid$.  More precisely, given a
conflicting (controller) action $a_c$, in an execution where
interaction $\alpha_i = a_c\mid a_i$ executes before $\alpha_j = a_c
\mid a_j$ for $j > i$, we can naturally replace $\mid h_{\alpha_i} -
h_{\alpha_j}\mid$ by $h_{\alpha_i} - h_{\alpha_j}$.  As for an
execution which violates this natural ordering (or ``canonicity''), we
show that we can make use of symmetry to rearrange it. First, we
formalise what we mean more precisely by \textit{canonicity}.  Given
an execution $\sigma$ and an interaction $\alpha_i = a_c\mid a_i$ we
denote by $\mpos(\sigma, \alpha_i)$ the last position of $\alpha_i$ in
$\sigma$. An execution $\sigma$ is canonical with respect to $a_c$ if
$\mpos(\sigma, \alpha_i) < \mpos(\sigma, \alpha_j)$ for any $i < j$.
Let $\exc$ be the set of canonical executions.  Thanks to symmetry,
any execution has a corresponding canonical execution. Assume $\sigma$
is such that there is a conflicting $a_c$ and for $i > j$ the last
occurrence of $\alpha_i = a_c\mid a_i$ appears latter than that of
$\alpha_j = a_c\mid a_j$. Let $\pi$ be such that $\pi(i) = j$ and
$\pi(j) = i$. Then $\pi(\sigma)$ is itself an execution and is
canonical.

For a canonical execution with $a_c$ being the action of interest
$\sep$ simplifies to:
\begin{align*}
  \sepc(\gamma) = \bigwedge_{\substack{i < j\\a_c \in \alpha_i \cap
      \alpha_j}} h_{\alpha_i} - h_{\alpha_j} \geq k_{a_c} \wedge
  \displaystyle{\bigwedge_{\substack{b \neq a_c\\ b \in \beta_i \cap \beta_j}}} \mid
  h_{\beta_i} - h_{\beta_j} \mid \geq k_b
\end{align*} 

We note that $\sepc$ reduces $\sep$ by $n!$. This is the best we can
get in general. However, under particular conditions, $\sep$ can be
further reduced. For instance, if the controller is such that it
considers components one by one and moreover, requires the use of some
designated action $a_c$, then $\sep$ further reduces to:
\begin{align*}
\displaystyle{\bigwedge_{a\in
    Act(C)}}\displaystyle{\bigwedge_{\substack{i < j\\a \in
      \alpha_i \cap \alpha_j}}} h_{\alpha_i} - h_{\alpha_j} \geq k_{a_c}
\end{align*}
This is because by considering components one by one, all conflicting
interactions involving the controller follow the same order as defined
for the designated action $a_c$. We anticipate and note that such a
scenario is the ``temperature controller'' case study from
Section~\ref{sec:impl}.

Finally, we show that for symmetric systems and properties it is
correct to consider $\sepc$ instead of $\sep$.
\begin{prop}
\label{prop:reds}
Let $M$ be a symmetric system, $\varphi$ be a symmetric property and
$\Phi$ the global invariant as defined in
Section~\ref{sec:methodEx}. We have that if $\vdash \Phi[\sep
\leftarrow \sepc] \rightarrow \varphi$ then $M \models \square
\varphi$.
\end{prop}
\proof (sketch) It suffices to show that $\vdash \Phi[\sep \leftarrow
\sepc] \rightarrow \varphi$ iff $\vdash \Phi
\rightarrow \varphi$. \\
``$\Leftarrow$'': trivial.  ``$\Rightarrow$'': It boils down to show
that if $\varphi$ is an invariant of $\exc$ then it is also an
invariant of the remaining executions $\sigma$ in $\exg \setminus
\exc$. If $\sigma$ does not have a conflicting action, we are done, as
$\sepc$ is an invariant by default. Else, we make use of the fact that
$\sigma$ has a canonical representation and that $\varphi$ is
symmetric.  \qed

An immediate application of the above reduction results in the
simplification we make use of in the temperature controller example
from Section~\ref{sec:impl}. Naturally, the results can be extended
also to systems with less symmetry by adapting the standard
constructions of automorphisms from, for example,
\cite{emersonS96}. More precisely, for a system $M$ for which $Aut(M)
= \{\pi \mid \pi(M) = M\}$ is a proper subgroup of $\Pi_n$, we need to
restrict to canonical executions which are consistent with the
permutations in $Aut(M)$. However, though such a generalisation is
possible, it is not clear if it is also useful: as it is well pointed
out in the literature about symmetries, determining $Aut(M)$ is, in
itself, a hard problem. This, together with the goal of keeping the
presentation as clear as possible, were the reasons why we strictly
considered only \textit{fully} symmetric systems.



\newcommand{\rodn}[1]{\mathit{Rod}_#1}

\section{Implementation and Experiments}
\label{sec:impl}

The method has been implemented in the RTD-Finder tool designed to
check safety properties for real-time component-based systems modelled
in the RT-BIP language \cite{rtbip11}.  The tool and the examples are
available at \url{http://www-verimag.imag.fr/RTD-Finder}.  

In RT-BIP, components are modelled as timed automata and synchronise
by means of n-ary multi-party interactions. The tool takes as input a
real-time BIP model and a file containing the safety property. It
subsequently generates a Yices \cite{yices} output file where the
invariants are expressed together with the property. RTD-Finder
proceeds by the following steps. It extends the components with
history clocks and computes their local invariants. The computation of
those invariants requires the implementation of several operations on
zones. For this purpose, we developed a DBM (Difference Bound
Matrices) library. RTD-Finder subsequently computes the history clocks
constraints and the interaction invariant. It writes all these
invariants to a file and calls Yices to check the satisfiabilty of $GI
\wedge \neg \Psi$.  If $GI \wedge \neg \Psi$ is unsatisfiable, the
property is valid. Otherwise, Yices generates a counter-example. We
note that, at present, the tool cannot conclude if it is a valid
counter-example, however, a guided backward analysis module is
currently under development. The benchmarks we used in our experiments
with RTD-Finder are described in what follows.
\subsection{Train gate controller (TGC)} 
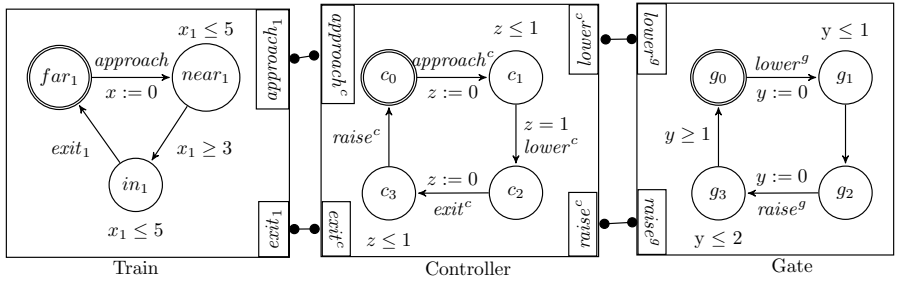
\begin{figure}[htp]
\centering
\input{tgc.tex}
\vspace*{-.2cm}
\caption{A controller interacting with a train and a gate}
\label{fig:tgc}
\vspace*{-.4cm}
\end{figure}

This is a classical example from \cite{alur94}.  The system is composed of
a controller, a gate and a number of trains.  For simplicity,
Figure~\ref{fig:tgc} depicts only one train interacting with the controller
and the gate. The controller lowers and raises the gate when a train
enters, respectively exits.
We propose to check  that when all the trains are at \textit{far} location, the gate cannot be going down ($g_2$ location).
The results are presented in Table.~\ref{tab:res}. 
When there are more than one train, be it $n$, the interactions $approach_i \mid approach$ (respectively $exit_i \mid
exit$), for $1 \geq i \geq n$ are in conflict on $approach$ (respectively
$exit$) of the controller.  In this case, in addition to the separation
constraints, we made use of the first heuristic presented in
Section~\ref{sec:h0inv}. More precisely, the invariant generated by the
heuristic is as follows:
\[
 \bigwedge_{i \neq j} \big( (h_{approach_i} \leq \te \wedge h_{approach_j} \leq \te) \rightarrow  h_{raise} \leq h_0\big) 
\]
\subsection{Fischer protocol} 
\begin{figure}[htp]
\centering
\input{fischer.tex}
\vspace*{-.2cm}
\caption{The Fischer protocol}
\label{fig:fischer}
\vspace*{-.4cm}
\end{figure}
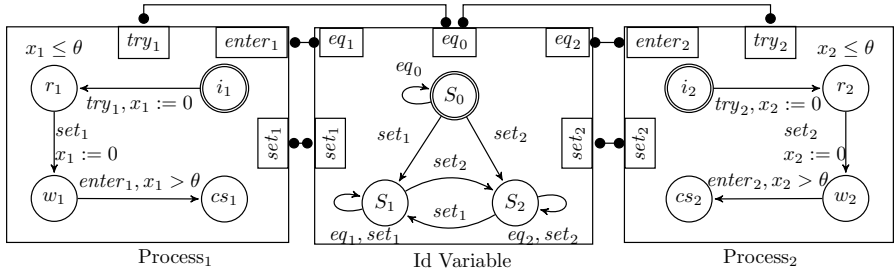

This is a well-studied protocol for mutual exclusion \cite{Lamport1987}.
The protocol specifies how processes can share a resource one at a time by
means of a shared variable to which each process assigns its own identifier
number. After $\theta$ time units, the process with the id stored in the
variable enters the critical state and uses the resource. We use an
auxiliary component \texttt{Id Variable} to mimic the role of the shared
variable.  The system with two concurrent processes is represented in
Figure~\ref{fig:fischer}. The property of interest is mutual exclusion:
$(cs_{i}\wedge {cs}_{j}) \rightarrow i=j$.

The component \texttt{Id Variable} has combinatorial behavior and a
large number of actions ($2n+1$), thus the generated invariant is huge
except for very small values of $n$. To overcome this issue, we made
use of the second heuristic presented in Section~\ref{sec:regex}.  To
simplify, we write $s_i$ instead of $\textit{set}_i$ and $e_i$ instead
of $\textit{eq}_i$. We construct the regular expression corresponding
to location $l_i$ and project it for actions $e_i,e_j,s_i,s_j$,
respectively $e_i,e_0,s_i,s_0$. The latter projection leads to the
following regular expression in restricted form:
\[ \mathbf{R_i}=(e_0+s_0)^*e_i. s_i+(e_0 +  s_0)^*s_i. e_i+(e_0 + e_i)^*s_0 s_i+(e_i + s_0)^*e_0 s_i+s_i \]
This regular expression translates into the following constraint on history clocks:
\begin{align*}
 \phi(\mathbf{R_i}) = & (h_{e_0} \geq h_{e_i} \wedge  h_{s_0} \geq h_{e_i}  \wedge h_{e_i} \geq h_{s_i} \wedge h_{e_i} \leq \te) \; {\vee}  \\
 & (h_{e_0} \geq h_{s_i} \wedge  h_{s_0} \geq h_{s_i}  \wedge h_{e_i} \leq h_{s_i} \wedge h_{s_i} \leq \te) \; {\vee}   \\
 & (h_{e_0} \geq h_{s_0}  \wedge  h_{e_i}  \geq h_{s_0}  \wedge  h_{s_0} \geq h_{s_i} \wedge h_{s_0} \leq \te) \; {\vee} \\
 & (h_{s_0} \geq h_{e_0}  \wedge  h_{e_i}  \geq h_{e_0}  \wedge  h_{e_0} \geq h_{s_i} \wedge h_{e_0} \leq \te) \;  {\vee} \\
 & ( h_{s_i} \leq \te \wedge h_{s_0}, h_{e_0}, h_{e_i} > \te)
\end{align*}
We deduce that $at(l_i) \rightarrow \phi(R_i)$ is an invariant of the
\texttt{Id Variable}, for any $i$.  These invariants in addition to
component invariants of processes and inequality constraints $\eqs(\gamma)$
are sufficient to show that mutual exclusion holds.
\subsection{Gear controller system} 
Our third example is taken from \cite{gear}. There it is described a
model of gear controller components in embedded systems operating
inside vehicles. A gear controller system is composed of five
components: an interface, a controller, a clutch, an engine and a
gear-box. The interface sends signals to the controller to change the
gear. In turn, the controller interacts with the engine, the clutch
and the gear-box. The engine is either regulating the torque or
synchronising the speed. The gear-box sets the gear between some fixed
bounds. The clutch works as the gear-box and it is used whenever the
engine is not able to function correctly (under difficult driving
conditions, for instance). One requirement that such a system should
satisfy in order to be correct is \textit{predictability}. This
requirement ensures a strict order between components. For instance,
it ensures that when the engine is regulating the torque, the clutch
is closed and the gear-box sets the gear. Another property of interest
that we checked is that the controller is in an error location only when
one of the other four components is in an error location also.
\subsection{Temperature controller (TC)}
\begin{figure}[htp]
\centering
\input{tc.tex}
\caption{A Controller interacting with two rods}
\label{fig:tc}
\vspace*{-.4cm}
\end{figure}
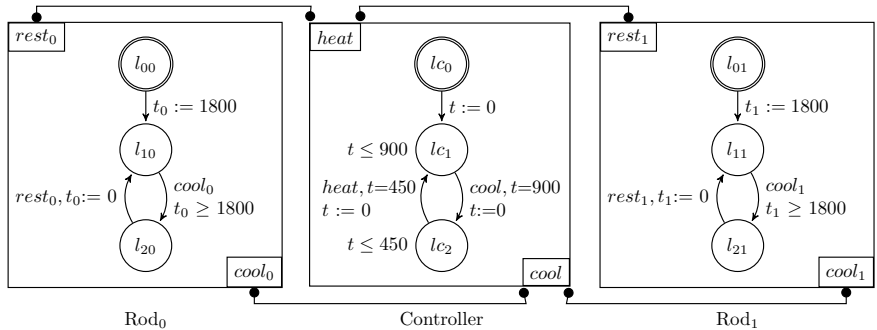

This example is an adaptation from
\cite{dfinder}. It represents a simplified model of a nuclear
plant. The system consists of a controller interacting with an
arbitrary number $n$ of rods (two, in Figure~\ref{fig:tc}) in order to
maintain the temperature between the bounds 450 and 900: when the
temperature in the reactor reaches 900 (resp. 450), a rod must be used
to cool (resp. heat) the reactor. The rods are enabled to cool only
after $900n$ units of time. The global property of interest is the
absence of deadlock, that is, the system can run continuously and keep
the temperature between the bounds. When the controller should take the \textit{cool} action,
at least one of the rods is ready to synchronise with it.
\mycomment{To express this property in our prototype, we adapt from \cite{tripakis99:progress} the definition of
\textit{enabled} states, while in Uppaal, we use the query \texttt{A[]
 not deadlock}. }
 For one rod, $\eqs(\gamma)$ is enough to show the
property. For more rods, because interactions are conflicting, we need
the separation constraints which basically bring as new information
conjunctions as $\wedge_i (h_{rest_{\pi(i)}} - h_{rest_{\pi(i-1)}}
\geq 1350)$ for $\pi$ an ordering on rods. Recalling the discussion
from Section~\ref{sec:sym}, such a reduction is correct because the
system enjoys the particularly helpful property of being symmetric.
\subsection{Dual chamber implantable pacemaker}
\begin{figure}[htp]
\centering
 \scalebox{.6}{
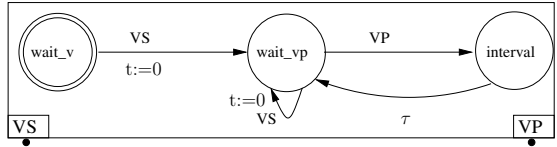
}
\scriptsize{
\caption{ Monitor for the upper rate limit property: the interval between a \textit{VS venticular event} and a \textit{VP venticular event} should be longer than \textit{TURI} \label{fig:pacemaker}}}
\end{figure}

As a last benchmark, we consider the verification of a dual chamber
implantable pacemaker presented in \cite{pacemaker}. A pacemaker is a
device for the management of the cardiac rhythm. It paces both the
atrium and the ventricle of the heart, and based on sensing both
chambers it can activate or inhibit further pacing. The model of
pacemakers we experimented with has five components, for (1) keeping
the heart rate above a minimum value, (2) maintaining delays between
atrial and ventricular activation, (3) preventing pacing the ventricle
too fast, filtering noise after (4) ventricular and (5) atrial events.
In our experiments, we considered the \textit{upper rate limit}
(\textit{URI}) property stating that the ventricles of the heart
should not be paced beyond a maximum rate, equal to a constant called
\textit{TURI}. The property states the existence of a minimum time
elapse\mycomment{, equal to a defined \textit{TURI} constant,} between
a ventricular sense (\textit{VS}) event and the following ventricular
pace (\textit{VP}) event. As in \cite{pacemaker}, we verified the
property by translating it into a monitor component which is shown in
Figure~\ref{fig:pacemaker}. The actions \textit{VS} and \textit{VP} of
the monitor are synchronised with those of the other components.
We verified that when the monitor reaches the location \textit{interval},
its clock $t$ is greater than \textit{TURI}. The corresponding property is 
\textit{interval} $\rightarrow t \geq$ \textit{TURI}.

Our method offers an additional way to check this property without
resorting to the monitor. We expressed it by means of the introduced
history clocks. The difference between the history clocks relative to
those two events is longer than the required time elapse:
\[(h_{VP} \leq h_{VS}  \wedge h_{VS} \leq h_0) \rightarrow  h_{VS} - h_{VP} \geq \textit{TURI}\]

\mycomment{ \q{ Should I put the following?}  Contrariwise, the method
  cannot verify that the time elapse between two consecutive
  \textit{VP venticular events} is bigger than $TURI$.  This is due to
  the fact that the history clocks record the last occurrence of each
  action. It results that in some way, the global invariant confounds
  consecutive occurrences of the same action. This is one of the cases
  where the method, being based on an over-approximation of the
  reachable states set, shows a false positive. A counter-example
  analysis module based on a guided backward analysis algorithm is
  under development and is meant to remedy this incompleteness.  }

\subsection{Results}
We ran our experiments on a Linux machine with Intel Core $3.20$ GHz
$\times 4$ and $15.6$ GiB memory. The results, synthesised in
Table~\ref{tab:res}, show the potential of our method in terms of
accuracy and scalability.  In Table~\ref{tab:res}, $n$ is the number
of components, $q$ is the total number of control locations, $c$
(resp. $h$) is the number of system clocks (resp. history clocks), $i$
is the number of interactions, while $t$ shows the total verification
time and $t_{yices}$ is the timed taken by Yices for satisfiability
checking of $GI \wedge \neg \Psi$. 


\mycomment{ As a side remark, we note
  that for each conflicting action $a$, the corresponding $k_a$
  constant, serving to express separation constraints, is computed
  while building the reachability graph of the containing
  component. The interaction inequalities for history clocks are
  necessary to catch the synchronization between the different
  components. However, the separation constraints are necessary only
  for verifying the temperature controller example.}
 \mycomment{Furthermore, the interaction invariant was not useful
  for any of the benchmarks described in this paper.  In fact, for
  most of the examples that we verified, history clocks constraints
  dispense RTD-Finder with their computation, since the safety
  properties could be verified without it. \\} \mycomment{ The
  heuristic based on symmetry exploitation (section 4.3 ) has not been
  implemented yet and is expected to reduce verification time for the
  temperature controller example, whereas the heuristic using regular
  expressions (section 4.2) is expected to compute efficiently the
  local invariant of the Id-Variable of Fischer protocol
\footnote{The heuristic for the
  automatic computation of the untimed Id-variable invariant for
  Fischer has not been implemented yet. For Fischer exclusively, $t$
  is the time required for the computation of all the other invariants
  and $q$ is the total number of locations of the other components.}. \\
}
\begin{table}[htp]
\centering{
\begin{tabular}{|c|c|c|c|c|c|c|c|}
\hline 
Model & $n$ & $q$  & $c$& $i$ & $h$ & $t$ & $t_{yices}$\tabularnewline
\hline 
\hline 
Train gate controller (50 trains) & 52 & 158 &  52 & 102 & 106 & 0.5s & 0.3s\tabularnewline
\hline 
Train gate controller (100 trains) & 102 & 308  & 102 & 202 & 206 & 5.3s & 0.6s\tabularnewline
\hline 
Train gate controller (200 trains) & 202 & 608  & 202 & 402 & 406 & 1m33s & 5s\tabularnewline
\hline 
Train gate controller (300 trains) & 302 & 908  & 302 & 602 & 606 & 9m8s & 20s \tabularnewline
\hline 
Train gate controller (500 trains) & 502 & 1508  & 502 & 1002 & 1006 & 1h13m20s & 2m52s \tabularnewline
\hline \hline 
Temperature controller (20 rods) & 21 & 42  & 21 & 40 & 42 & 0.07s & 0.01s \tabularnewline
\hline 
Temperature controller (50 rods) & 51 & 102  & 51 & 100 & 102 & 0.35s & 0.04s \tabularnewline
\hline 
Temperature controller (100 rods) & 101 & 204 &  102 & 200 & 204 & 3.7s & 0.08s\tabularnewline
\hline 
Temperature controller (300 rods) & 301 & 602 &  302 & 600 & 602 & 5m47s & 0.9s\tabularnewline
\hline \hline 
Fischer protocol (100 processes) & 101 & 400 & 101  & 300 & 501 & 2.7s & 0.06s \tabularnewline
\hline 
Fischer protocol (200 processes) & 201 & 800 & 201  & 600 & 1001 & 0m47s & 0.22s\tabularnewline
\hline 
Fischer protocol (300 processes) & 301 & 1200 &  301 & 900 & 1501 & 4m27s & 0.5s \tabularnewline
\hline \hline
Gear controller & 5 & 65 & 4 & 17 & 32 & 15.1s & 0.14s \tabularnewline
\hline \hline 
Pacemaker (with monitor) & 7 & 19 &  11 & 6 & 21 & 15.23s & 0.044s\tabularnewline
\hline 
Pacemaker (without monitor) & 6 & 16 &  9 & 6 & 19 & 15s & 0.032s\tabularnewline
\hline 
\end{tabular}
\caption{ Results from experiments} 
\label{tab:res}
}
\end{table}

To the best of our knowledge, there are no tools to compositionally
verify safety properties of timed systems. Consequently, there are no
relevant tools to compare RTD-Finder with. Netherveless, we did a
small comparison with Uppaal \cite{uppaal}. Uppaal is a well-known model-checking
tool which is highly optimised. For instance, thanks to some reduction
techniques, it has better scores on the first example (the TGC system)
in particular and on smaller systems in general.  Nonetheless,
generally, state space exploration is costly. This can be
illustrated by means of the temperature controller example: for
$10$ rods, Uppaal generated no results after five hours and $436519$
explored states. On the other hand, RTD-Finder checked the property for
$300$ rods in few minutes, as shown in Table~\ref{tab:res}.  The
timings for the RTD-Finder tool are obtained by the java command
\texttt{getCpuTime} called to compute the total verification time,
while the results for Uppaal come from the command \texttt{verifyta}
which comes with the Uppaal 4.1.14 distribution.


\section*{Related Work} 
Automatic generation of invariants for concurrent systems is a
long-time studied topic.  Yet, to our knowledge, specific extensions
or applications for timed systems are rather limited. As an exception,
the papers \cite{badban10:inv-ta,fietzke12:inv-ta} propose a
monolithic, non-compositional method for finding invariants in the
case of systems represented as a single timed automaton.

Compositional verification for timed systems has been mainly
considered in the context of timed interface theories \cite{Alfaro02}
and contract-based assume guarantee reasoning
\cite{Ecdar,AutomaticCompERAs,mocha}. These methods usually rely upon
choosing a ``good'' decomposition structure and require individual
abstractions for components to be deterministic timed I/O
automata. Finding the abstractions is in general difficult, however,
their construction can be automated by using learning techniques
\cite{AutomaticCompERAs} in some cases. In contrast to the above, we
are proposing a fully automated method generating, in a compositional
manner, an invariant approximating the reachable states of a timed
system.  

Abstractions serve also for compositional minimisation, for instance
\cite{berendsen08} minimises by constructing timed automata quotients
with respect to simulation; these quotients are in turn composed for
model-checking. Our approach is orthogonal in that we do not compose
at all. Compositional deductive verification as in \cite{boer97} is
also orthogonal on our work in that, by choosing a particular class of
local invariants to work with, we need not focus on elaborate proof
systems but reason at a level closer to intuition.

The use of additional clocks has been considered, for instance, in
\cite{Bengtsson98, pettersson07:partial}. There, extra reference
clocks are added to components to faithfully implement a partial order
reduction strategy for symbolic state space exploration. Time is
allowed to progress desynchronised for individual components and
re-synchronised only when needed, i.e., for direct interaction within
components. Clearly, the history clocks in our work behave in a
similar way, however, our use of clocks is as a helper construction in
the generation of invariants and we totally avoid global state space
exploration. Finally, another successful application of extra clocks
has been provided in \cite{SalahBM09} for timing analysis of
asynchronous circuits. There, specific history clocks are reset on
input signals and used to provide a new time basis for the
construction of an abstract model of output signals of the circuit.

\mycomment{
\todo{abstraction + 3 refs from the last reviewer }

Finally, we note that there is notable work on abstraction \cite{...}.
This work is orthogonal on our approach: the abstractions are with
respect to global states, consequently, their applicability to systems
with a great number of components is problematic.

For \cite{tiwari11:hybrid} underline that they do k-induction or
bounded model-checking, that is, the goal is to invalidate safety
properties \obs{not quite sure because they analyse why k-ind doesn't
  return safe for some workbenches}, while in our case, the goal is to
validate: if no solution, then we can conclude that the system is
safe. In addition, the approach in \cite{tiwari11:hybrid} relies on
quantifier elimination which does only works for a small number of
quantifiers (while for large systems this condition cannot be
satisfied/enforced/guaranteed).

\cite{tiwari05:inv-hybrid}

\cite{pettersson07:partial}
}

\section{Conclusions}
\label{sec:conc}
We presented a fully automated compositional method to generate global
invariants for timed systems described as parallel compositions of
timed automata components using multi-party interactions. The
soundness of the method proposed has been proven. In addition, it has
been successfully tested on several benchmarks.
This method has been implemented in the RTD-Finder tool. The results
show that it may outperform the existing exhaustive
exploration-based techniques for large systems, thanks to the use of
compositionality and over-approximations.
Nonetheless, the generated invariant is an over-approximation of the 
reachable states set and false-positives may raise. To remedy this, 
we are working on a guided backward analysis module to decide upon 
their validity. \\
 In order to achieve a better integration, we are working on handling 
 richer classes of systems, including systems with data variables 
 and {\it urgencies} \cite{BozgaSifakis06} on transitions.
 Actually, urgencies provide an alternative way to constrain time progress, which is more intuitive to
use by programmers but very difficult to handle in a compositional
way. A second direction of research which is potentially interesting
for systems containing identical, replicated components and closely
related to the symmetry-based reduction is the application of our
method to the verification of parameterised timed systems. Finally, we
are considering specific extensions to particular classes of timed
systems and properties, in particular, for schedulability analysis of
systems with mixed-critical tasks.

\mycomment{
We presented a fully automated compositional method to generate global
invariants for timed systems described as parallel compositions of
timed automata components using multi-party interactions. The
soundness of the method proposed has been proven. In addition, it has
been implemented and successfully tested on several benchmarks.  The
results show that our method may outperform existing exhaustive
exploration-based techniques for large systems, thanks to the use of
compositionality and over-approximations.

This work is currently being extended in several directions.  First,
we work on integrating it within D-Finder tool \cite{dfinder} and the
Real-Time BIP framework \cite{rtbip11}.  In order to achieve a better
integration, we are working on handling {\it urgencies}
\cite{BozgaSifakis06} on transitions.  Actually, urgencies provide an
alternative way to constrain time progress, which is more intuitive to
use by programmers but very difficult to handle in a compositional
way. A second direction of research which is potentially interesting
for systems containing identical, replicated components and closely
related to the symmetry-based reduction is the application of our
method to the verification of parameterised timed systems. Finally, we
are considering specific extensions to particular classes of timed
systems and properties, in particular, for schedulability analysis of
systems with mixed-critical tasks.}

\subsection*{Acknowledgement.} We are grateful to the anonymous
referees for their constructive input and for their thorough feedback.
We would also like to thank our colleague Mahieddine Dellabani for his
help with two benchmarks.

\bibliographystyle{abbrv} 
\bibliography{main}

\end{document}

%% file: commonCmds.tex
\newcommand{\mycomment}[1]{}

\newcommand{\todo}[1]{\textcolor{OrangeRed}{TODO: #1}}

\newcommand{\q}[1]{\textcolor{Fuchsia}{Q: #1}}

\newcommand{\mor}[1]{\displaystyle{\bigvee_{#1}}}

\newcommand{\hp}{\mathcal{H}_A}
\newcommand{\ha}{\mathcal{H}_{\gamma}}
\newcommand{\actions}{\mathit{Act}}
\newcommand{\hist}{\mathcal{H}}

\newcommand{\eqs}{\mathcal{E}}
\newcommand{\eqsc}{\mathcal{E}^*}

\newcommand{\X}{\mathcal{X}}

\newcommand{\en}{\mathit{enabled}}

\newcommand{\msucc}{\mathsf{succ}}
\newcommand{\dsucc}{\mathsf{disc}\_\mathsf{succ}}
\newcommand{\tsucc}{\mathsf{time}\_\mathsf{succ}}

\newcommand{\inv}{\mathsf{tpc}}
\newcommand{\close}{\mathsf{norm}}
\newcommand{\bv}{\bold{v}}

\newcommand{\transit}[1]{\stackrel{#1}{\rightarrow}}
\newcommand{\reach}{\mathit{Reach}}

\newcommand{\cn}{\mathit{B}\xspace}

\newcommand{\ctn}{\mathit{Controller}}
\newcommand{\cwk}{\mathit{Worker}}

\newcommand{\cwkO}{\mathit{Worker}_1}

\newcommand{\true}{\mathit{true}\xspace}

\newcommand{\ic}{\mathit{CI}}
\newcommand{\iim}{\mathit{II}}

\newcommand{\sep}{\mathcal{S}}

\newcommand{\vr}{(VR)}

\newcommand{\te}{h_0}

\newcommand{\compgamma}{B^*}
\newcommand{\sysgammah}{\compgamma \|_{\gamma^h} B_i^h}

\newcommand{\ctrl}{C}

\newcommand{\dl}{l^*}

\newcommand{\exg}{Exec}

\newcommand{\exc}{Exec^{c}}

\newcommand{\sepc}{\sep^{c}}

\newcommand{\mpos}{\mathit{lpos}}

\graphicspath{{Figs/}}

%% file: tikzEx.tex
\newcommand{\rode}[3]{%
  \node[state, accepting] (r0#1) at(#2,#3) {\small $l_{0 #1}$};%
  \node[state] (r1#1) [below=.6cm of r0#1] {\small $l_{1 #1}$};%
  \node[state] (r2#1) [below=.8cm of r1#1] {\small $l_{2 #1}$};%
  \path%
  (r0#1) edge node[right]{\parbox{2cm}{\small $t_{#1}:=1800$}}
  (r1#1)%
  (r1#1) edge[bend left] node[right]{\parbox{2.3cm}{\small $cool_{#1}\\ t_{#1}$ $\geq$ $1800$}} (r2#1)%
  (r2#1) edge[bend left] node[left]{\parbox{1.9cm}{\small $rest_{#1},
      t_{#1}$:= 0}} (r1#1);%
\node[draw,left=1.5cm of r0#1.north] (ar#1) {$rest_{#1}$};%
\node[draw,right=1.5cm of r2#1.south] (ac#1) {$cool_{#1}$};%
\node[draw, rectangle, inner sep=0, fit=(ar#1) (ac#1)] {};%
\node[below=.6cm of r2#1.south] (R#1) {Rod$_{#1}$};%
}

\newcommand{\malpha}{900}
\newcommand{\mlambda}{450}

\newcommand{\controller}[2]{%
  \node[state, accepting] (AC) at(#1,#2) {$lc_0$};%
  \node[state] (BC) [below=.6cm of AC] {$lc_1$};%
  \node[left=.0cm of BC.west] {\small $t \leq \malpha$};%
  \node[state] (CC) [below=.8cm of BC] {$lc_2$};%
  \node[left=.0cm of CC.west] (invC2) {\small $t \leq \mlambda$};%
  \node[below=.6cm of CC.south] {Controller};%
  \path%
  (AC) edge node[right]{\parbox{1.cm}{\small$t:=0$}} (BC)%
  (BC) edge[bend left] node[right]{\parbox{1.9cm}{\small $cool, t$=$\malpha$\\$t$:=0}} (CC)%
  (CC) edge[bend left] node[left]{\parbox{1.7cm}{\small $heat,t$=$\mlambda$ \\ $t$ := 0}} (BC);%
drawing ports
\node[draw,left=1.5cm of AC.north] (ch) {$heat$};%
\node[draw,right=1.5cm of CC.south] (cc) {$cool$};%
bounding box for controller
\node[draw, rectangle, inner sep=0, fit=(ch) (cc)] {};%
}

\newcommand{\cTikz}[2]{%
\node[state, accepting] at (#1,#2) (l0) {\small $lc_0$};%
\node[state]         (l1) [below=.7cm of l0] {\small $lc_1$};%
\node [left=.05cm of l1] {\small $x \leq 4$};%
\node[state]         (l2) [below=.8 of l1] {\small $lc_2$};%
\path%
(l0) edge node [right] {\small \parbox{1.3cm}{$x \geq 4n$\\ $x := 0$}} (l1)%
(l1) edge[bend left] node[right] (gb) {\parbox{1.3cm}{\small $a, x = 4$\\$x$:=0}} (l2)%
(l2) edge[bend left] node[left] (gd) {\parbox{.95cm}{\small $c$\\$x:=0$}} (l1);%
\node[draw,right=1.1cm of l0.north east] (a) {$a$};%
\node[draw,right=1.15cm of l2.south east] (c) {$c$};%
bounding box for controller
\node[draw, rectangle, inner sep=0, fit= (gd) (a) (c)] (boxc) {};%
\node[left=.3cm of boxc.north,xshift=.2cm,yshift=.3cm] (ctrl) {$\ctn$};
}

\newcommand{\wTikz}[3]{%
\node[state, accepting] (l3#1) at (#2,#3) {\small $l^{#1}_1$};%
\node[state]         (l4#1) [below=.7cm of l3#1] {\small $l^{#1}_2$};%
\path%
(l3#1) edge[bend left] node[right] (gb#1) {\parbox{1.35cm}{\small $b_{#1}, 
                                                                   y_1 \geq 8$}} (l4#1)%
(l4#1) edge[bend left] node[left] (gd#1) {\parbox{1.46cm}{\small $d_{#1}, 
                                                                  y_1:=0$}} (l3#1);%
\node[draw,left=1.2cm of l3#1.north west] (b#1) {$b_{#1}$};%
\node[draw,left=1.2cm of l4#1.south west] (d#1) {$d_{#1}$};%
\node[draw, rectangle, inner sep=0, fit=(b#1) (d#1) (gb#1) (gd#1)] (boxw#1) {};%
\node[below=.2cm of boxw#1.south] (w#1) {$\cwk_{#1}$};%
}

\newcommand{\wTikzR}[3]{%
\node[state, accepting] (l3#1) at (#2,#3) {\small $l^{#1}_1$};%
\node[state]         (l4#1) [below=.7cm of l3#1] {\small $l^{#1}_2$};%
\path%
(l3#1) edge[bend left] node[right] (gb#1) {\parbox{1.35cm}{\small $b_{#1},
                                                                   y_2  \geq 8$}} (l4#1)%
(l4#1) edge[bend left] node[left] (gd#1) {\parbox{1.46cm}{\small $d_{#1}, 
                                                                   y_2:=0$}} (l3#1);%
\node[draw,right=1.1cm of l3#1.north east] (b#1) {$b_{#1}$};%
\node[draw,right=1.1cm of l4#1.south east] (d#1) {$d_{#1}$};%
\node[draw, rectangle, inner sep=0, fit=(b#1) (d#1) (gd#1) (gb#1)] (boxw#1) {};%
\node[below=.2cm of boxw#1.south] (w#1) {$\cwk_{#1}$};%
}

\newcommand{\wTikzO}[4]{%
\tikzset{draw opacity=#4}
\node[state, accepting,opacity=#4*#4] (l3#1) at (#2,#3) {\small $l_{1 #1}$};%
\node[state,opacity=#4*#4]         (l4#1) [below=.8cm of l3#1] {\small $l_{2 #1}$};%
\path%
(l3#1) edge[bend left,opacity=#4*#4] node[right] (gb#1) {\parbox{1.2cm}{\small $b_{#1}\\y_{#1} \geq 4n$}} (l4#1)%
(l4#1) edge[bend left,opacity=#4*#4] node[left] (gd#1) {\parbox{1.1cm}{\small $d_{#1}\\y_{#1}:=0$}} (l3#1);%
\node[draw,left=.8cm of l3#1.north west,opacity=#4] (b#1) {$b_{#1}$};%
\node[draw,left=.8cm of l4#1.south west,opacity=#4] (d#1) {$d_{#1}$};%
\node[draw, rectangle, inner sep=0, fit=(b#1) (d#1) (gb#1) (gd#1), opacity=1*#4] (boxw#1) {};%
\node[right=.3cm of boxw#1.north, xshift=.2cm, yshift=.2cm, opacity=#4] (w#1) {$\cwk_{#1}$};%
}

\newcommand{\chTikz}[2]{%
\node[state, accepting] at (#1,#2) (l0) {\small $lc_0$};%
\node[state]         (l1) [below=.7cm of l0] {\small $lc_1$};%
\node [left=.05cm of l1] {\small $x \leq 8$};%
\node[state]         (l2) [below of=l1] {\small $lc_2$};%
\path%
(l0) edge node [right] {\small $x\geq 4$ \\ $x := 0$} (l1)%
(l1) edge[bend left] node[right] (gb) {\parbox{1.3cm}{\small $a, x = 8$\\$x$:=0\\$\bf{h_{a}:=0}$}} (l2)%
(l2) edge[bend left] node[left] (gd) {\parbox{1.3cm}{\small $c$\\$x:=0$\\$\bf{h_{c}:=0}$}} (l1);%
\node[draw,right=1.cm of l0.north east] (a) {$a$};%
\node[draw,right=1.cm of l2.south east] (c) {$c$};%
bounding box for controller
\node[draw, rectangle, inner sep=0, fit= (gd) (a) (c)] (boxc) {};%
\node[below=.3cm of boxc.south] (ctrl) {$\ctn^h$};
}

\newcommand{\whTikz}[3]{%
\node[state, accepting] (l3#1) at (#2,#3) {\small $l^{#1}_1$};%
\node[state]         (l4#1) [below of=l3#1] {\small $l^{#1}_2$};%
\path%
(l3#1) edge[bend left] node[right] (gb#1) {\parbox{1.35cm}{\small $b_{#1}, y \geq 8$\\$\bf{h_{b_{#1}}:=0}$}} (l4#1)%
(l4#1) edge[bend left] node[left] (gd#1) {\parbox{1.2cm}{\small $d_{#1}$\\$y:=0$\\$\bf{h_{d_{#1}}:=0}$}} (l3#1);%
\node[draw,left=1.1cm of l3#1.north west] (b#1) {$b_{#1}$};%
\node[draw,left=1.1cm of l4#1.south west] (d#1) {$d_{#1}$};%
\node[draw, rectangle, inner sep=0, fit=(b#1) (d#1) (gd#1) (gb#1)] (boxw#1) {};%
\node[below=.5cm of boxw#1.south] (w#1) {$\cwk_{#1}^h$};%
}

\newcommand{\whTikzR}[3]{%
\node[state, accepting] (l3#1) at (#2,#3) {\small $l^{#1}_1$};%
\node[state]         (l4#1) [below of=l3#1] {\small $l^{#1}_2$};%
\path%
(l3#1) edge[bend left] node[right] (gb#1) {\parbox{1.35cm}{\small $b_{#1}, y \geq 8$\\$\bf{h_{b_{#1}}:=0}$}} (l4#1)%
(l4#1) edge[bend left] node[left] (gd#1) {\parbox{1.2cm}{\small $d_{#1}$\\$y:=0$\\$\bf{h_{d_{#1}}:=0}$}} (l3#1);%
\node[draw,right=1.2cm of l3#1.north east] (b#1) {$b_{#1}$};%
\node[draw,right=1.2cm of l4#1.south east] (d#1) {$d_{#1}$};%
\node[draw, rectangle, inner sep=0, fit=(b#1) (d#1) (gd#1) (gb#1)] (boxw#1) {};%
\node[below=.5cm of boxw#1.south] (w#1) {$\cwk_{#1}^h$};%
}

\newcommand{\chTikzSym}[2]{%
\node[state, accepting] at (#1,#2) (l0) {\small $lc_0$};%
\node[state]         (l1) [below=.7cm of l0] {\small $lc_1$};%
\node [left=.05cm of l1] {\small $x \leq 4$};%
\node[state]         (l2) [below of=l1] {\small $lc_2$};%
\path%
(l0) edge node [right] {\small \parbox{1.3cm}{$x \geq 8$\\$x := 0$}} (l1)%
(l1) edge[bend left] node[right] (gb) {\parbox{1.3cm}{\small $a, x = 4$\\$x$:=0\\$\bf{h_{a}:=0}$}} (l2)%
(l2) edge[bend left] node[left] (gd) {\parbox{1.3cm}{\small $c$\\$x:=0$\\$\bf{h_{c}:=0}$}} (l1);%
\node[draw,right=1.cm of l0.north east] (ar) {$a$};%
\node[draw,right=1.cm of l2.south east] (cr) {$c$};%
\node[draw,left=1.2cm of l0.north west] (al) {$a$};%
\node[draw,left=1.2cm of l2.south west] (cl) {$c$};%
bounding box for controller
\node[draw, rectangle, inner sep=0, fit= (gd) (al) (cl) (ar) (cr)] (boxc) {};%
\node[below=.3cm of boxc.south] (ctrl) {$\ctn^h$};
}

\newcommand{\trainH}[2]{%
\begin{tikzpicture}[->,>=stealth',shorten >=1pt,auto,node distance=4.cm,semithick]%
\tikzstyle{every state}=[fill=white,text=black]%
  \node[state, accepting] (AT) {\small far};%
  \node[above=.4cm of AT] (Train) {Train$^h$};%
  \node[above=.03cm of AT] (auxt) {};%
  \node[state]         (BT) [right of=AT] {\small near};%
  \node[right=.1cm of BT] {\small x $\le$ 5};%
  \node[state]         (CT) [below of=BT] {\small in};%
  \node[right=.1cm of CT] {\small x $\le$ 5};%
 \path%
        (AT) edge              node[above]{approach} node[below=.5cm]{x:= 0, hat:=0} (BT)%
        (BT) edge              node{($x > 2$)?} (CT)%
        (CT) edge              node{\parbox{1cm}{exit\\het:=0}} (AT);%
\end{tikzpicture}%
}

\newcommand{\aapproach}{\mathit{approach}}
\newcommand{\aexit}{\mathit{exit}}

\newcommand{\train}[2]{%
  \node[state, accepting] (AT) at (#1, #2) {\small $far_{1}$};
  \node[left=.1cm of AT] (auxt) {};
  \node[state]         (BT) [right=1.45cm of AT] {\small $near_{1}$};
  \node[right=.8cm of AT] (x) {}; 
  \node[above=1.2cm of BT.south] {\small $x_1 \le 5$};
  \node[state]         (CT) [below=1.45cm of x.west] {\small $in_{1}$};
  \node[below=.06cm of CT] (inv) {\small $x_1 \le 5$};
  \node[below=.13cm of inv] (Train) {Train};
  \path 
        (AT) edge              node[above]{$\aapproach$} node[below]{\small $x:= 0$} (BT)
       (BT) edge              node{$x_1 \geq 3$} (CT)
        (CT) edge              node{$\aexit_1$} (AT);
  \node[draw, rotate=90, right=1.2cm of BT.south,xshift=.04cm] (at) {$\aapproach_1$};
  \node[draw, rotate=90, right=2.46cm of inv.south, xshift=-.2cm,yshift=-.03cm] (et) {$\aexit_1$}; 
\node[draw, rectangle, inner sep=0, fit=(at) (et) (auxt) (BT) (CT) (AT) (inv)] {};
}

\newcommand{\araise}{\mathit{raise}}
\newcommand{\alower}{\mathit{lower}}

\newcommand{\gate}[2]{%
  \node[state, accepting] (AG)  at (#1, #2) {\small $g_0$};
  \node[state]         (BG) [right=1.3cm of AG] {\small $g_1$};
  \node[right=.1cm of BG] (auxg) {};
  \node[above=.05cm of BG] (inv1) {\small y $\leq$ 1};
  \node[state]         (CG) [below=1.1cm of BG] {\small  $g_2$};
  \node[state]         (DG) [below=1.1cm of AG] {\small  $g_3$};
  \node[below=.05cm of DG] (inv2) {\small y $\le$ 2};
  \node[right=.7cm of DG] (xg) {}; 
 \node[below=.9cm of xg] (Gate) {Gate};
  \node[draw, left=.7cm of AG,rotate=-90,xshift=-.05cm] (lg) {$\alower^g$};
  \node[draw, left=1.2cm of inv2.south,rotate=-90] (rg) {$\araise^g$};
  \path 
        (AG) edge              node[above]{$\alower^g$} node[below]{\small $y := 0$} (BG)
        (BG) edge              (CG)
        (CG) edge              node[below]{$\araise^g$} node[above]{\small $y := 0$} (DG)
        (DG) edge              node[left]{\small $y \geq 1$ } (AG);
\node[draw, rectangle, inner sep=0, fit=(lg) (rg) (auxg) (BG)] {};
}

\newcommand{\controllerTrain}[2]{%
  \node[state, accepting] (AC) at (#1,#2) {$c_0$};
  \node[state] (BC) [right=1.3cm of AC] {$c_1$};
  \node[above=.1cm of BC] {\small $z \leq 1$};
  \node[state] (CC) [below=1.1cm of BC] {$c_2$};
  \node[state] (DC) [below=1.1cm of AC] {$c_3$};
  \node[right=.8cm of DC] (xc) {}; 
  \node[below=.95cm of xc] {Controller};
  \node[below=.1cm of DC] (inv3) {\small $z \leq 1$};
  \node[draw, left=.95cm of inv3.south,rotate=-90] (ec) {$\aexit^c$};
  \node[draw, right=.63cm of BC,rotate=90,xshift=.05cm,yshift=-.1cm] (lc) {$\alower^c$};
  \node[draw, left=.92cm of AC.south,rotate=-90,xshift=-.02cm] (ac) {$\aapproach^c$};
  \node[draw, right=3.5cm of inv3.south,rotate=90] (rc) {$\araise^c$};
  \path 
  (AC) edge node[above]{$\aapproach^c$} node[below]{\small $z := 0$} (BC)
  (BC) edge node[right]{\parbox{2cm}{\small $z=1$\\ $\alower^c$}} (CC)
  (CC) edge node[above]{\small $z := 0$} node[below]{$\aexit^c$} (DC)
  (DC) edge node[left]{$\araise^c$} (AC);
\node[draw, rectangle, inner sep=0, fit=(ec) (ac) (rc) (lc)] {};
}

\newcommand{\aset}{\mathit{set}}
\newcommand{\aeq}{\mathit{eq}}

\newcommand{\fischerV}[2]{%
  \node[state] (AV) at (#1, #2) {\small $S_1$};
  \node[below=.03cm of AV] (auxv) {};
  \node[state]         (BV) [right=1.45cm of AV] {\small $S_2$};
  \node[right=.8cm of AV] (x) {}; 
  \node[state, accepting]  (CV) [above=1.45cm of x.west] {\small $S_0$};
  \node[below=.6cm of x] (IdV) {Id Variable};
  \path 
        (AV) edge[loop left]              node[below=.3cm, xshift=.4cm]{\parbox{1cm}{$\aeq_1, \aset_1$}} (AV)
        (BV) edge[loop right]              node[below=.3cm, xshift=-.4cm]{{$\aeq_2,\aset_2$}} (BV)
        (CV) edge[loop left]              node[above=.2cm, xshift=.1cm]{{$\aeq_0$}} (CV)
        (CV) edge              node[left,yshift=.25cm]{$\aset_1$} (AV)
        (CV) edge              node{{$\aset_2$}} (BV)
        (AV) edge[in=0,out=10,bend left]  node[above]{{$\aset_2$}} (BV)
        (BV) edge[in=0,out=10,bend left]  node[above]{{$\aset_1$}} (AV);
  \node[draw, above=.25cm of CV.north] (eq0) {$\aeq_0$};
  \node[draw, right=1.2cm of eq0] (eq2) {$\aeq_2$};
  \node[draw, below=1.5cm of eq2, anchor=north, xshift=-.11cm,rotate=90] (set2) {$\aset_2$};
  \node[draw, left=1.2cm of eq0] (eq1) {$\aeq_1$};
  \node[draw, left=4.3cm of set2, anchor=west, rotate=90] (set1) {$\aset_1$}; 
\node[draw, rectangle, inner sep=0, fit=(set1) (set2) (eq0) (eq1) (eq2) (auxv) (AV) (CV) (BV)] {};
}

\newcommand{\aenter}{\mathit{enter}}
\newcommand{\atry}{\mathit{try}}

\newcommand{\fischerP}[3]{%
  \node[state, accepting] (AP)  at (#1, #2) {\small $i_{#3}$};
  \node[state]         (BP) [right=(#3*3-5)*1.9cm of AP] {\small $r_{#3}$};
  \node[right=.1cm of BP] (auxp) {};
  \node[above=-.cm of BP] (inv#3) {\small $x_{#3} \leq \theta$};
  \node[state]         (CP#3) [below=1.1cm of BP] {\small  $w_{#3}$};
  \node[state]         (DP) [below=1.1cm of AP] {\small  $cs_{#3}$};
  \node[right=(#3*3-5)*.7cm of DP] (xp) {}; 
  \node[below=.6cm of xp] (Process$_{#3}$) {Process$_{#3}$};
  \path 
        (AP) edge              node[below=(#3*.02-.02)]{$\atry_{#3}, x_{#3} := 0$} (BP)
        (BP) edge              node[left=(#3*1.1-2.55)]{\parbox{1.3cm}{\small $\aset_{#3}$\\$x_{#3} := 0$}} (CP#3)
        (CP#3) edge              node[above=(#3*.02-.02)]{$\aenter_{#3},x_{#3} > \theta$}(DP);
}

\newcommand{\fischerL}[2]{
\fischerP{#1}{#2}{1}
  \node[draw, left=.55cm of eq1, anchor=east] (enter1) {$\aenter_1$};
  \node[draw, left=.8cm of enter1] (try1) {$\atry_1$};
  \node[draw, left=1.cm of set1, anchor=west, rotate=90] (set11) {$\aset_1$};
  \node[below=.2cm of CP1.south west, xshift=-.4cm] (auxpl) {};
  \node[draw, rectangle, inner sep=0, fit=(enter1) (set11) (inv1) (auxpl)] {};
}

\newcommand{\fischerR}[2]{
\fischerP{#1}{#2}{2}
  \node[draw, right=1.9cm of eq2, anchor=east] (enter2) {$\aenter_2$};
  \node[draw, right=.8cm of enter2] (try2) {$\atry_2$};
  \node[draw, right=1.1cm of set2, anchor=east, rotate=90] (set22) {$\aset_2$};
  \node[below=.2cm of CP2.south east, xshift=.4cm] (auxpr) {};
  \node[draw, rectangle, inner sep=0, fit=(enter2) (set22) (inv2) (auxpr)] {};
}

\newcommand{\figregexp}[2]{%
\node[state, accepting] (l0) at (#1,#2) {\small $l_0$};
\node[state, right of=l0] (l1) {\small $l_1$};
\path 
     (l0) edge[loop above] node[above] (a) {$a$} (l0)
     (l0) edge[bend left] node[above]{$b$} (l1)
     (l1) edge[bend left] node[above] (b) {$b$} (l0)
     (l1) edge[loop above] node[above] (c) {$c$} (l1);
\node[draw, below=.65cm of l0.south west] (la) {$a$};
\node[draw, below=.5cm of b] (lb) {$b$};
\node[draw, right=2.45cm of la] (lc) {$c$};
\node[draw, rectangle, inner sep=0, fit=(l0) (la) (a) (c) (l1) (lc)] {};
}

\newcommand{\figregexph}[2]{%
\node[state, accepting] (l0) at (#1,#2) {\small $l_0$};
\node[state, right of=l0] (l1) {\small $l_1$};
\path 
     (l0) edge[loop above] node[above] (a) {$a,h_a:=0$} (l0)
     (l0) edge[bend left] node[above]{$b,h_b:=0$} (l1)
     (l1) edge[bend left] node[below] (b) {$b,h_b:=0$} (l0)
     (l1) edge[loop above] node[above] (c) {$c,h_c:=0$} (l1);
\node[draw, below=.65cm of l0.south west, xshift=-.3cm] (la) {$a$};
\node[draw, below=-.09cm of b] (lb) {$b$};
\node[draw, right=3.05cm of la] (lc) {$c$};
\node[draw, rectangle, inner sep=0, fit=(l0) (la) (a) (c) (l1) (lc)] {};
}

%% file: abstractExTikz.tex
\resizebox{9.cm}{!}{
\begin{tikzpicture}[->,>=stealth',shorten >=1pt,auto,node distance=2.cm,semithick,font=\small]
\tikzstyle{every state}=[fill=white,text=black, inner sep=0]%
\cTikz{0}{0}
\foreach \o [count=\oi from 1] in {1, 0.35, 0.1} {
\wTikzO{\oi}{4.5+\o}{.06-\o}{\o};
  \draw[*-*,opacity=\o] (a.east) -- (b\oi.west);
  \draw[*-*,opacity=\o] (c.east) -- (d\oi.west);
}
\end{tikzpicture}
}


%% file: regexpTikz.tex
\resizebox{10.cm}{!}{
\begin{tikzpicture}[->,>=stealth',shorten >=1pt,auto,node distance=2.2cm,semithick,font=\small]
\tikzstyle{every state}=[fill=white,text=black, inner sep=0]%
\figregexp{0}{0}
\figregexph{8}{0}
\end{tikzpicture}
}


%% file: tgc.tex
\resizebox{12.cm}{!}{
\begin{tikzpicture}[->,>=stealth',shorten >=1pt,auto,node distance=4.cm,semithick,font=\small]
\tikzstyle{every state}=[fill=white,text=black]

 \train{0}{0}
 \controllerTrain{5.95}{0}
 \gate{11.9}{0}

  \draw[*-*] (at.south) -- (ac.south);
  \draw[*-*] (et.south) -- (ec.south);
  \draw[*-*] (lg.south) -- (lc.south);
  \draw[*-*] (rg.south) -- (rc.south);

\end{tikzpicture}
}


%% file: fischer.tex
\resizebox{12.cm}{!}{
\begin{tikzpicture}[->,>=stealth',shorten >=1pt,auto,node distance=4.cm,semithick,font=\small]
\tikzstyle{every state}=[fill=white,text=black,inner sep=0, minimum size=.8cm]

\fischerV{6}{-2}
\fischerL{3.2}{0}{1}
\fischerR{11.3}{0}{2}

  \draw[*-*] (set22.north) -- (set2.south);
  \draw[*-*] (eq2.east) -- (enter2.west);
  \draw[*-*] (eq0.north)++(.15,0) |- ++(.,.4) -| (try2.north);
  \draw[*-*] (set11.south) -- (set1.north);
  \draw[*-*] (eq1.west) -- (enter1.east);
  \draw[*-*] (eq0.north)++(-.15,0) |- ++(.,.4) -| (try1.north);

\end{tikzpicture}
}


%% file: tc.tex
\resizebox{12.cm}{!}{
\begin{tikzpicture}[->,>=stealth',shorten >=1pt,auto,node distance=4.cm,semithick,font=\small]
\tikzstyle{every state}=[fill=white,text=black]

 \rode{0}{0}{0}
 \controller{5.5}{0}
 \rode{1}{11}{0}

  \draw[*-*] (ar0.north) -- +(., .3) -- +(5.1,.3) -- (ch.north west);
  \draw[*-*] (ar1.north) -- +(-.,.3) -- +(-5., .3) -- (ch.north east);
  \draw[*-*] (ac0.south) -- +(., -.3) -- +(5.05, -.3) -- (cc.south west);
  \draw[*-*] (ac1.south) -- +(-., -.3) -- +(-5.2, -.3) -- (cc.south east);
\end{tikzpicture}
}


%% file: Figs/pacemaker.pdf_t
\begin{picture}(0,0)%
\includegraphics{pacemaker.pdf}%
\end{picture}%
\setlength{\unitlength}{4144sp}%
\begingroup\makeatletter\ifx\SetFigFont\undefined%
\gdef\SetFigFont#1#2#3#4#5{%
  \reset@font\fontsize{#1}{#2pt}%
  \fontfamily{#3}\fontseries{#4}\fontshape{#5}%
  \selectfont}%
\fi\endgroup%
\begin{picture}(5469,1448)(1384,-1677)
\put(2566,-961){\makebox(0,0)[lb]{\smash{{\SetFigFont{12}{14.4}{\rmdefault}{\mddefault}{\updefault}{\color[rgb]{0,0,0}t:=0}%
}}}}
\put(5311,-1456){\makebox(0,0)[lb]{\smash{{\SetFigFont{12}{14.4}{\rmdefault}{\mddefault}{\updefault}{\color[rgb]{0,0,0}$\tau$}%
}}}}
\put(3601,-1276){\makebox(0,0)[lb]{\smash{{\SetFigFont{12}{14.4}{\rmdefault}{\mddefault}{\updefault}{\color[rgb]{0,0,0}t:=0}%
}}}}
\end{picture}%